\documentclass[twocolumn, superscriptaddress, nofootinbib]{revtex4-1}
\usepackage{amsmath}
\usepackage[english]{babel}
\usepackage[pdftex]{graphicx}
\usepackage{color}

\usepackage[sort&compress]{natbib}
\usepackage{hyperref}
\graphicspath{{images/}{./}}

\usepackage{amsmath}
\usepackage{amssymb}
\usepackage{mathtools}
\usepackage{braket}

\renewcommand{\vec}[1]{\ensuremath{\boldsymbol{#1}}}

\newcommand{\plane}[2]{\vec\Phi_{#1,#2}}

\renewcommand{\deg}{^{\circ}}

\newcommand{\eee}{{\cal E}}
\newcommand{\sgn}{\,\mbox{\rm sgn}}
\newcommand{\diag}{\,\mbox{\rm diag}}
\begin{document}
\title{Anomalous Floquet tunneling in uniaxially strained graphene}

\author{Yonatan Betancur-Ocampo}
\email{ybetancur@icf.unam.mx}
\affiliation{Instituto de Ciencias F\'isicas, Universidad Nacional Aut\'onoma de M\'exico, Cuernavaca 62210, M\'exico}

\author{Parisa Majari}
\email{majari@icf.unam.mx}
\affiliation{Instituto de Ciencias F\'isicas, Universidad Nacional Aut\'onoma de M\'exico, Cuernavaca 62210, M\'exico}

\author{Diego Espitia}
\email{despitia@icf.unam.mx}
\affiliation{Instituto de Ciencias F\'isicas, Universidad Nacional Aut\'onoma de M\'exico, Cuernavaca 62210, M\'exico}

\author{Fran\c{c}ois Leyvraz}
\email{leyvraz@icf.unam.mx}
\affiliation{Instituto de Ciencias F\'isicas, Universidad Nacional Aut\'onoma de M\'exico, Cuernavaca 62210, M\'exico}
\affiliation{Centro Internacional de Ciencias, Cuernavaca 62210, M\'exico}

\author{Thomas Stegmann}
\email{stegmann@icf.unam.mx}
\affiliation{Instituto de Ciencias F\'isicas, Universidad Nacional Aut\'onoma de M\'exico, Cuernavaca 62210, M\'exico}

\begin{abstract}
The interplay of strain engineering and photon-assisted tunneling of electrons in graphene is considered 
for giving rise to atypical transport phenomena. The combination of uniaxial strain and 
a time-periodic potential barrier helps to control the particle transmission for a wide range of tunable 
parameters. With the use of the tight-biding approach, the elasticity theory, and the Floquet scattering,
 we found an angular shift of the maximum transmission in the sidebands for uniaxial strains breaking the mirror symmetry with respect to the normal incidence, which is called anomalous Floquet tunneling. We 
 show that electron tunneling depends strongly on the barrier width, incident angle, uniaxial 
 strain, and the tuning of the time-periodic potential parameters. An adequate modulation of the barrier 
 width and oscillation amplitude serves to select the transmission in the sidebands. These findings can 
 be useful for controlling the electron current through the photon-assisted tunneling being used 
 in multiple nanotechnological applications. 
\end{abstract}

\maketitle

\section{Introduction}

Photon-assisted tunneling is a powerful tool for controlling electron current in a device through the illumination of a particular area of the system \cite{Dayem1962, Hartman1962, Tien1963, Shirley1965, Buettiker1982, Fletcher1985, Grossmann1991, Wagner1994, Kouwenhoven1994, Wagner1995, Blick1995, Wagner1996, Wagner1997, Platero2004, Zhang2006, Trauzettel2007, Schwede2010, Mueller2010, Freitag2012, Tielrooij2013, Ma2018, Gudmundsson2012, Biswas2013, Soltani2020}. The tuning of the laser frequency and the intensity can serve to explore different features in quantum transport. The understanding of the interaction of electrons under external electromagnetic fields has led to a huge number of technological applications. Nevertheless, there are many unusual electronic transport effects in the presence of time-periodic potentials that requires an exhaustive study and revision from the new perspective given by the rising of two-dimensional materials \cite{Neto2009, Vogt2012, Li2014b, Wehling2014, Carvalho, Li2018b, Betancur-Ocampo2019, Betancur-Ocampo2020, Biswas2019}. Most of these materials belong to the classification of Dirac matter, where the Dirac-Weyl equation describes the dynamics of low energy excitations \cite{Neto2009, Wehling2014, Diaz-Bautista2020, Setare2019, Diaz-Bautista2019, DellAnna2018, Majari2017, Yang2019, Ren2019, Ghosh2019, Le2020}. While that, the Floquet scattering formalism has been the most recurrent theory for depicting the dynamics of photon-assisted tunneling \cite{Li1999, Moskalets2002, Gu2011, Bilitewski2015, Savelev2011}. This approach allows a simplified vision of electron tunneling through sidebands. Electrons impinging the oscillating potential barrier are reflected or refracted from different energy channels by the absorption or emission of one or multiple photons \cite{Li1999, Wurl2019, Zeb2008, Cao2011, Sattari2020, Jongchotinon2020, Yan2017, Jellal2014, Savelev2012, Szabo2013, Chen2015, Schulz2015}. In this way, Floquet scattering has been used successfully for explaining the constructive interference of continuum and bound states in quantum wells, an effect known as Fano resonances \cite{Fano1961, Goeres2000, Kobayashi2002, Miroshnichenko2010, Lu2012, Myoung2013, Zhang2015}. Other interesting phenomena have been predicted based on the Floquet scattering theory, among them the Wigner delay times \cite{Sattari2020}, Hartman effect \cite{Kh2018}, suppression of Klein tunneling \cite{Zeb2008, Biswas2013, Sinha2012, Yampolextquotesingleskii2008}, Floquet topological insulators \cite{Calvo2011, Usaj2014, Perez-Piskunow2015,  Savelev2011, Cayssol2013, Wintersperger2020, Wang2013, Mahmood2016, Afzal2020}, non-Hermitian Floquet invisibility \cite{Longhi2017}, and photo-electronic induced emission \cite{Liang2013, Sun2012, Tao2020}.

The interplay between strain engineering and photon-assisted tunneling may open more possibilities, due to the increment of external variables to control the electron tunneling. By applying strain in graphene and related materials, the electronic band structure is modified drastically and serves to modulate the electronic, optical, and transport properties \cite{Pereira2009, Pereira2009a, Pereira2010, Ribeiro2009, Cocco2010, Pellegrino2010, Choi2010, Naumis, Naumov2011, Rostami2012, Barraza-Lopez2013, Assili2015, Guinea2009, Levy2010, Haddad2018, Sahalianov2018, Contreras-Astorga2020, Zahidi2020, Concha2018, Lee2015, Midtvedt2016, Perez-Pedraza2020}. Inhomogeneous strain gave rise to the emergence of valleytronics and pseudo-magnetic fields \cite{Guinea2009, Levy2010, Wu2011, Rechtsman2012, Stegmann2016, Stegmann2019}. Outstanding electron optics-like effects appear in uniaxially strained graphene \cite{Betancur-Ocampo2018, Diaz-Bautista2020}. Such a system displayed partial positive refraction in asymmetric Veselago lenses, a negative reflection of electrons, and anomalous Klein tunneling \cite{Betancur-Ocampo2018, Zhang2019, Betancur-Ocampo2019}. Those theoretical results may be tested not only in uniaxially strained graphene but also hexagonal optical lattices and photonic crystals \cite{Wintersperger2020, Tarruell2012, Rechtsman2012}. Recently, a time periodic potential in optical lattices was experimentally realized in \cite{Wintersperger2020}. Photonic crystal emulations of strained graphene evidenced that Klein tunneling persists for deformations along the zig-zag and armchair directions \cite{Bahat-Treidel2010}. 

In this paper, we show that the combination of photon-assisted tunneling and strain engineering present singular transmission effects. The application of uniaxial strain causes a strong anisotropy in the electron tunneling. Dependent on the amplitude of time-periodic potential and frequency, there are preferential incidence angles for electron tunneling in the sidebands. We find suppression of the anomalous Klein tunneling, which can be useful for electronic confinement. Moreover, the tuning of potential barrier width serves to select the transmission in sidebands in order to produce photo-induced electronic current.   

The paper is structured as follows: In section \ref{model}, we give a short review of how the tight-binding approach and elasticity theory is useful for the development of a strain-modified Hamiltonian in graphene and related anisotropic hexagonal lattices. In section \ref{Floquet}, we apply the Floquet scattering theory in order to analyze the transmission features of a fully strained graphene sheet with time-periodic potential barrier. We present in \ref{Discussion} the results of our numerical and analytical calculations of the transmission probabilities for the sidebands. We expose the conclusions and final remarks in section \ref{Conclusions}.

\section{Dirac-Weyl Hamiltonian of uniaxially strained graphene}\label{model}

\begin{figure*}[t!!]
  \begin{tabular}{cc}
  (a) \qquad \qquad \qquad \qquad \qquad \qquad \qquad \qquad \qquad & (b) \qquad \qquad \qquad \qquad \qquad \qquad \qquad \qquad \qquad \qquad \\ 
  \includegraphics[width=0.47\textwidth]{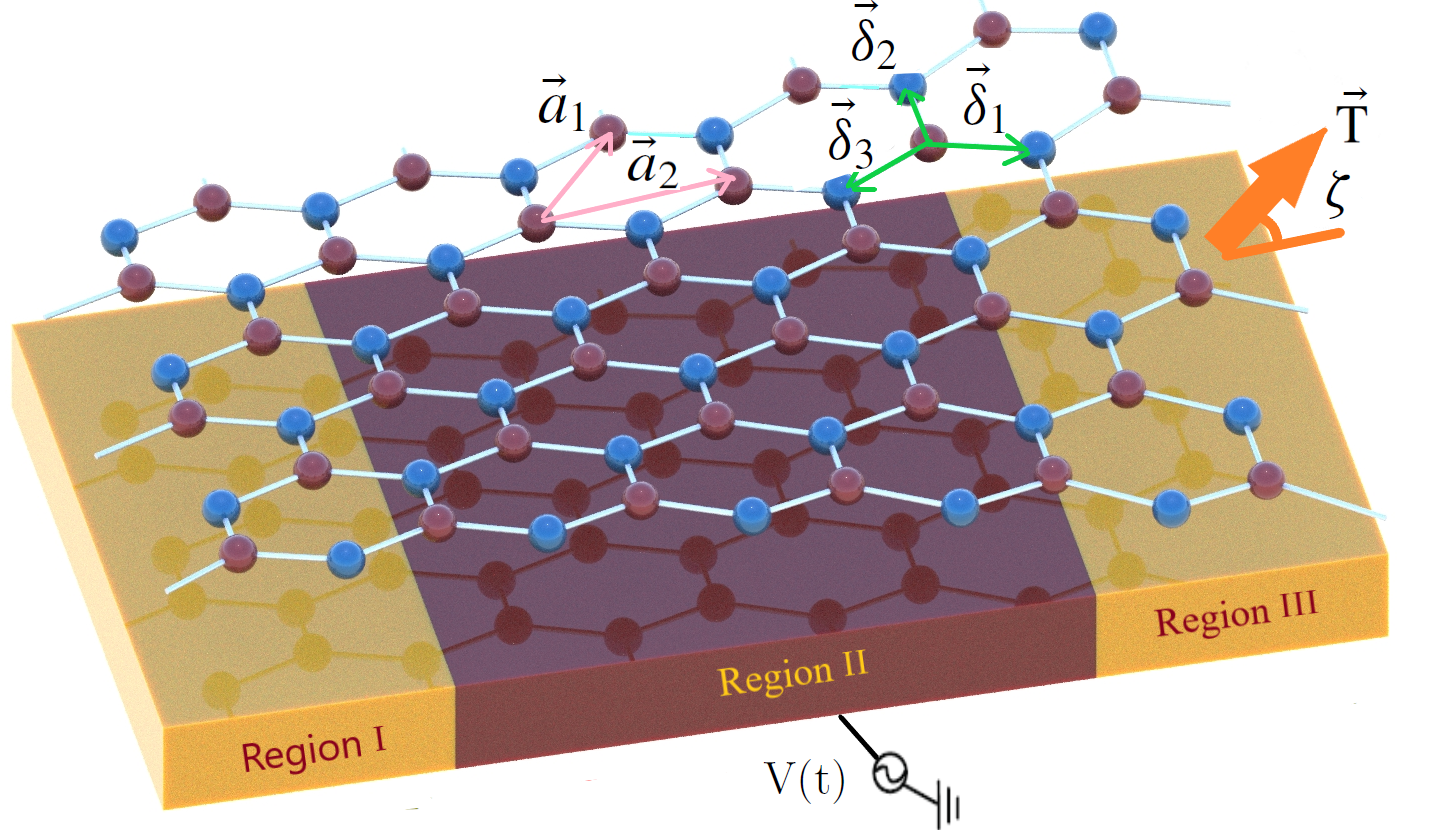} &
  \includegraphics[width=0.49\textwidth]{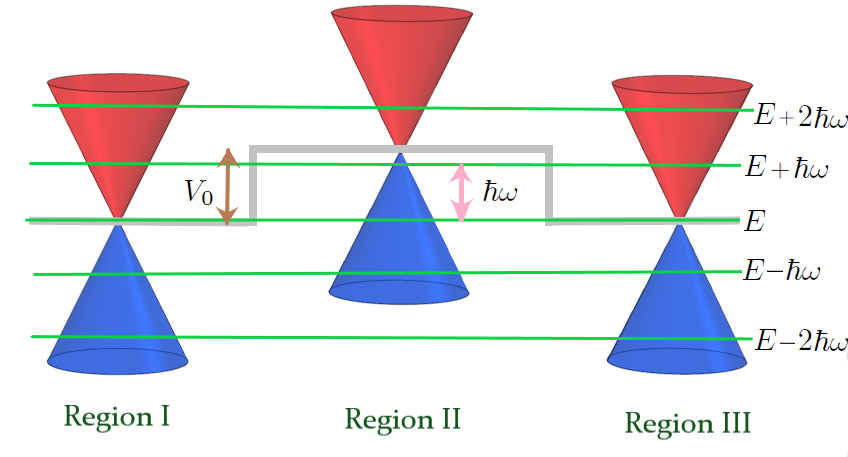}
  \end{tabular}
  \caption{(a) Schematic representation of uniaxially strained graphene in a time-periodic potential. Violet and blue circles indicate the sites of triangular sublattices A and B, respectively. The vector $\vec{T}$ corresponds to the applied tension in the $\zeta$ direction. Each nearest-neighbor possesses a hopping parameter $\tau_j$ and bond length $\delta_j$. The quantities $\vec{a}_1$ and $\vec{a}_2$ are the deformed lattice vectors. The external gates induce a time-periodic potential barrier in region II. (b) Description of the Floquet scattering across the time-periodic potential and Dirac cone structure. Horizontal green lines represent the energy channels $E - m \hbar\omega$, where transmission in the sidebands occur.}
  \label{fig1}
\end{figure*}

Uniaxially strained graphene and anisotropic hexagonal lattices are composed by two deformed triangular sublattices $A$ and $B$ with a basis of two atoms per unit cell, as shown in Fig.~\ref{fig1}(a). According to the elasticity theory, the application of a uniaxial strain deformes the lattice vectors in the pristine configuration, and they are given by \cite{Pereira2009, Betancur-Ocampo2018, Diaz-Bautista2020}

\begin{eqnarray}
\vec{a}_1 = (a_{1x}, a_{1y}) & = & \sqrt{3}a(1 + \rho^{-}\epsilon + \rho^{+}\epsilon\cos2\zeta, \rho^{+}\epsilon\sin2\zeta) \nonumber\\
 \vec{a}_2 = (a_{2x}, a_{2y})& = & \frac{\sqrt{3}}{2}a(1 + \rho^{-}\epsilon + 2\rho^{+}\epsilon\cos(2\zeta -60\deg), \nonumber\\
 & & \sqrt{3}(1 + \rho^{-}\epsilon) + 2\rho^{+}\epsilon\sin(2\zeta -60\deg)), \nonumber\\
 &&
\label{lattvs}
\end{eqnarray}

\noindent where the constants $\rho^{\pm}$ are defined as 

\begin{equation}
\rho^{\pm} = \frac{1}{2}(1 \pm \nu)
\end{equation}
\noindent and $\nu=0.18$ is the Poisson ratio, while $a = 0.142$ nm is the bond length in pristine graphene \cite{Neto2009}. The vectors $\vec{\delta}_j$ with $j = 1, 2$ and $3$ indicate the three nearest neighbors site on the underlying sublattice $A$, as shown in Fig. \ref{fig1}(a). The strain parameters $\epsilon$ and $\zeta$ quantify the percentage of tensile strain and the direction of the applied tension $\vec T$ with respect to the $x$ axis. The failure strain has been estimated to occur at the approximated value $\epsilon \approx 28 \%$ \cite{Cadelano2009,Lee2008}. However, we use a moderated range of $\epsilon$ from 0 to 10$\%$ in all our calculations within the linear elastic regimen, where Tight-Binding (TB) and Density Functional Theory (DFT) calculations have been demonstrated to have a good agreement \cite{Ribeiro2009}. Nevertheless, controlled and reversible extreme strains $\epsilon > 10 \%$ has been realized experimentally \cite{Garza2014}. Using the TB approach to first 
nearest neighbors, we consider one orbital per atom in the unit cell and neglect the overlap orbital among neighboring sites. 
The scaling rule $\tau_j=\tau \exp[-\beta(\delta_j/a-1)]$ relates the hopping parameters $\tau_j$ with the deformed bond lengths 
$\delta_j$. In graphene, $\beta = 2.6$ is the Grüneisen constant and $\tau = 2.7$ eV is the isotropic hopping \cite{Wong2012, Neto2009}. 
This scaling rule evidences that the Fermi velocity is anisotropic and has a tensorial character. In the Fourier basis and 
expanding around the Dirac cone, the Hamiltonian is \cite{Betancur-Ocampo2018, Diaz-Bautista2020}

\begin{equation}\label{HD}
{{H}_D}= \hbar\begin{bmatrix}
0& \vec{v}^{c*}\cdot\vec{k}\\
  \vec{v}^c\cdot\vec{k} & 0
\end{bmatrix},
\end{equation}

\noindent where $\vec{k} = (k_x, k_y)$ is the wave vector and 

\begin{equation}\label{Complvs}
\vec{v}^c = (v^c_x,v^c_y) = \frac{i}{\hbar}\left(\vec{a}_1\tau_1{e}^{-i\vec{K}_D\cdot\vec{\delta}_1}
 + \vec{a}_{2}\tau_2{e}^{-i\vec{K}_D\cdot\vec{\delta}_2}\right),
\end{equation}

\noindent are the complex velocities, being $\vec{K}_D$ the Dirac point position, which is the solution of
\begin{equation}\label{DPs}
\sum^3_{j=1}\tau_j{e}^{-i\vec{K}_D\cdot\vec{\delta}_j} = 0.
\end{equation}
The Hamiltonian \eqref{HD} is the Dirac-Weyl type $H = v_{ij}\sigma_i p_j$, where $v_{ij}$ is the 
Fermi velocity tensor and $p_j = \hbar k_j$ are the components of linear momentum. The electronic band structure 
of anisotropic hexagonal lattice, in the semimetallic phase, present generally elliptical and rotated 
Dirac cones. The dispersion relation
\begin{equation}
E = s\hbar|\vec{v}^c\cdot\vec{k}|
\label{eq:disp}
\end{equation}
of the Hamiltonian \eqref{HD} displays this cone around the Dirac point in the reciprocal space, where $s$ is the 
band index \cite{Betancur-Ocampo2018}.  The eigenstates of the Dirac-Weyl-like Hamiltonian \eqref{HD} 
have the form  
\begin{equation}
\plane{s}{\vec k}(\vec r)=\frac{1}{\sqrt{2}}\binom{1}{s e^{i \phi(\vec k)}}e^{i\vec k\cdot\vec r}
\label{eq:defphi}
\end{equation}
and the definition of pseudospin angle $\phi(\vec k)$ is
\begin{subequations}
\begin{eqnarray}
\tan\phi(\vec k) &=& \frac{-v_xk_x\sin\mu_x + v_yk_y\sin\mu_y}{v_xk_x\cos\mu_x + v_yk_y\cos\mu_y},\\
v^c_x&=&v_xe^{-i \mu_x},\\
v^c_y&=&v_ye^{i \mu_y}.
\end{eqnarray}
\label{defphi}
\end{subequations}
\noindent Here $v_{x,y}^{c}$ are the $x$- and $y$-components of the complex vector $\vec v^{c}$ with norm $v_{x,y}$ and phase $\mu_{x,y}$. 
The pseudo-spin direction, wave vector, and the group velocity are generally not parallel. 

The direction of group velocity is found to be \cite{Betancur-Ocampo2018}
\begin{equation}\label{theta}
\tan\theta = \frac{v^2_yk_y + v_xv_yk_x\cos(\mu_x + \mu_y)}{v^2_xk_x + v_xv_yk_y\cos(\mu_x + \mu_y)}
\end{equation}
and allows to obtain  the wave vector in terms of incidence angle. The application of uniaxial strains out of the 
zig-zag and armchair direction have led to the emergence of anomalous Klein tunneling \cite{Betancur-Ocampo2018}, which occurs 
at the incidence angle 
\begin{equation}\label{KTA}
\theta_{KT} = \arctan[v_y\cos(\mu_x + \mu_y)/v_x]
\end{equation}
when $k_y = 0$ in Eq. \eqref{theta}.  

We now rewrite the dispersion relation (\ref{eq:disp})
in the more explicit form
\begin{eqnarray}
E&=s \hbar\sqrt{k_x^2 v_x^2+2 k_y k_x
v_x^2 \tan\theta_{KT}+k_y^2 v_y^2}.
\label{eq:explE}
\end{eqnarray}

In next sections, we shall evidence that this symmetry breaking with respect to the $x$ axis modifies drastically the 
electron transmission for the sidebands. 

\section{Photon-assisted tunneling through a time-periodic potential barrier}\label{Floquet}

 We study the tunneling of electrons in uniaxially strained graphene under the presence of a time-periodic potential 
 barrier, as shown in Fig. \ref{fig1}. The photon-assisted mechanism, such as a time-periodic potential used here, causes the appearance of many sidebands \cite{Li1999, Zeb2008, Sinha2012}. These sidebands correspond to multiple copies of the dispersion relation with a relative energy separation $\hbar\omega$, where $\hbar$ and $\omega$ is the Planck constant and the potential frequency, respectively. The Floquet scattering is the usual theory to describe the tunneling of a single electron with energy $E$ to cross the time-periodic potential gaining or losing the energy quantity $m\hbar\omega$, where $m = 0, \pm 1, \pm 2, \ldots$ indicates the sideband (see Fig. \ref{fig1}(b)). The tunneling is elastic (inelastic) if the electron crosses the oscillating barrier without (with) changes in the energy. Most of the experimental realizations that involved photon-assisted tunneling are observed generally in the frequency range from the microwave to infrared electromagnetic spectrum \cite{Dayem1962, Blick1995, Kouwenhoven1994, Platero2004, Schwede2010, Mueller2010, Freitag2012, Tielrooij2013, Ma2018, Sun2012, Wang2013, Mahmood2016, Soltani2020, Tao2020}.
 
 We have several external variables to control the electron tunneling by means of the application of uniaxial strain and tuning of amplitude, frequency, barrier height and width of the time-periodic potential. From a general point to view, we write the time-dependent Schr\"odinger equation as 

\begin{equation}\label{TDWE}
[H(\vec{p}) +V(\vec{r},t)]\psi(\vec{r},t)=i\hbar \partial_t \psi(\vec{r},t),
\end{equation}
\noindent where $H(\vec{p})$ can be a general Hamiltonian that depends only on the linear momentum $\vec{p}$. Therefore, the following development from Eqs. \eqref{TDWE} to \eqref{eq:defkappa} can be applied to multiple systems in condensed matter to depict the Floquet scattering of electrons in the presence of a time-periodic potential barrier.

The eigenvectors of $H(\vec{p})$ are the wave functions $\plane{s}{\vec k}(\vec{r})$ 
of the electron belonging to the momentum $\vec k$ and band index $s$. For instance, in the particular Hamiltonian \eqref{HD} the wave functions $\plane{s}{\vec k}(\vec{r})$ are given by \eqref{eq:defphi}. We define $\eee(s,\vec k)$ to be the corresponding eigenvalue
\begin{equation}
H(\vec{p})\plane{s}{\vec k}(\vec r)=\eee(s,\vec k)\plane{s}{\vec k}(\vec r).
\end{equation}
The time-periodic potential is given by
\begin{equation}\label{TPP}
V(x,t) =
\begin{cases}
 V_0+V_1 \cos(\omega t),  \; \; \textrm{for} \; \; {0}<x< {D} \\
0, \; \; \textrm{otherwise}
\end{cases}
\end{equation}
To solve this, we divide the system in three regions, namely $x\leq0$, $0\leq x\leq D$ and $D\leq x$, denoted by I, II, and III respectively. 
We define
\begin{equation}
\alpha=\frac{V_1}{\hbar\omega}
\end{equation}
and find the general plane wave solutions for all three regions: 
\begin{subequations}
\begin{eqnarray}
W_{s_{\text{I}}\vec k_{\text{I}}}(\vec r, t)&=&\plane{s_{\text{I}}}{\vec k_{\text{I}}}(\vec r)\exp\left[
-i\eee(s_{\text{I}},\vec k_{\text{I}})t/\hbar
\right]
\label{eq:defWa}
\\
W_{s_{\text{II}}\vec k_{\text{II}}}(\vec r, t)&=&\plane{s_{\text{II}}}{\vec k_{\text{II}}}
(\vec r)\exp\left[
-(i/\hbar)
\big(
V_0t\right.\nonumber\\
&& \qquad \left.+\eee(s_{\text{II}},\vec k_{\text{II}})t+V_1\sin(\omega t)/\omega
\big)
\right]\nonumber\\
&=&\plane{s_{\text{II}}}{\vec k_{\text{II}}}(\vec r)\sum_{m=-\infty}^\infty J_m(\alpha)
\exp\left[
-(i/\hbar)
\big(
V_0\right.\nonumber\\
&& \qquad \left.+\eee(s_{\text{II}},\vec k_{\text{II}})+m\hbar\omega
\big)
t\right]
\label{eq:defWb}
\\
W_{s_{\text{III}}\vec k_{\text{III}}}(\vec r, t)&=&\plane{s_{\text{III}}}{\vec k_{\text{III}}}(\vec r)\exp\left[
-i\eee(s_{\text{III}},\vec k_{\text{III}})t/\hbar
\right]
\label{eq:defWc}
\end{eqnarray}
\label{eq:defW}
\end{subequations}

\noindent The second equality in (\ref{eq:defWb}) follows from the identity
\begin{eqnarray}
\exp\left[
-i\alpha\sin(\omega t)
\right]&=&\sum_{m=-\infty}^\infty J_m(\alpha) e^{-im\omega t},
\end{eqnarray}
\noindent where $J_m(\alpha)$ are Bessel functions of the first kind. We now determine linear superpositions of these various solutions $\plane{s_{\text{I}}}{\vec k_{\text{I}}}(\vec r, t)$, $\plane{s_{\text{II}}}{\vec k_{\text{II}}}(\vec r, t)$,
and $\plane{s_{\text{III}}}{\vec k_{\text{III}}}(\vec r, t)$ in such a way as  
to yield continuous behaviour at the interfaces $x=0$ 
and $x=D$ for all times. 

We assume that the incoming wave in region I is characterized by a wave vector $\vec k_0$ 
and a band index $s_0$. To match this in region II, we need all momenta $\vec{q}_m$ such that
\begin{subequations}
\begin{eqnarray}
\label{SBs}
E&:=&\eee(s_0,\vec k_0)=\eee(s_m^\prime,\vec q_m^{\pm})+V_0+m\hbar\omega\\
k_{y,0}&=&q_{y,m}^{\pm},
\end{eqnarray}
\label{eq:defkappa}
\end{subequations}

\noindent where the second relation follows from the conservation of $k_{y,0}$ at all interfaces. The expressions \eqref{SBs} are the sidebands for the system described by the Hamiltonian $H(\vec p)$ in the presence of the time-periodic potential $V(x,t)$. Using the specific Hamiltonian of uniaxially strained graphene \eqref{HD} in this general development, we have 
\begin{subequations}
\begin{eqnarray}\label{kpr}
 q_{m,x}^{\pm}& = & \pm s'_m\sqrt{\frac{(E-V_0-m\hbar\omega)^2}{ 
\hbar^2v_x^2}-{\frac{v^2_y}{v^2_x}k_{y,0} ^2\sin^2(\mu_x+\mu_y)}} \nonumber\\
& & -k_{y,0}\tan\theta_{KT},\\
s_m^\prime&=&\sgn(E-V_0 -m\hbar\omega),
\end{eqnarray}
\end{subequations}
\noindent where the $\pm$ sign indicates the two possible solutions for $q_m$, which are obtained by the dispersion relation \eqref{eq:explE}. There are two states with these wave vectors $q_m$ and the same energy, we shall use
them to represent ``left-going'' and ``right-going'' waves in region II, in the same way as it would happen
with $k_x$ and $-k_x$ in isotropic systems. 
\begin{figure*}[t]
\begin{tabular}{ccc}
(a) \qquad \qquad \qquad \qquad \qquad \qquad \qquad \qquad& (b) \qquad \qquad \qquad \qquad \qquad \qquad 
\qquad & (c) \qquad \qquad \qquad \qquad \qquad \qquad \qquad\\
\includegraphics[trim = 1mm 0mm 12mm 0mm, scale= 0.39, clip]{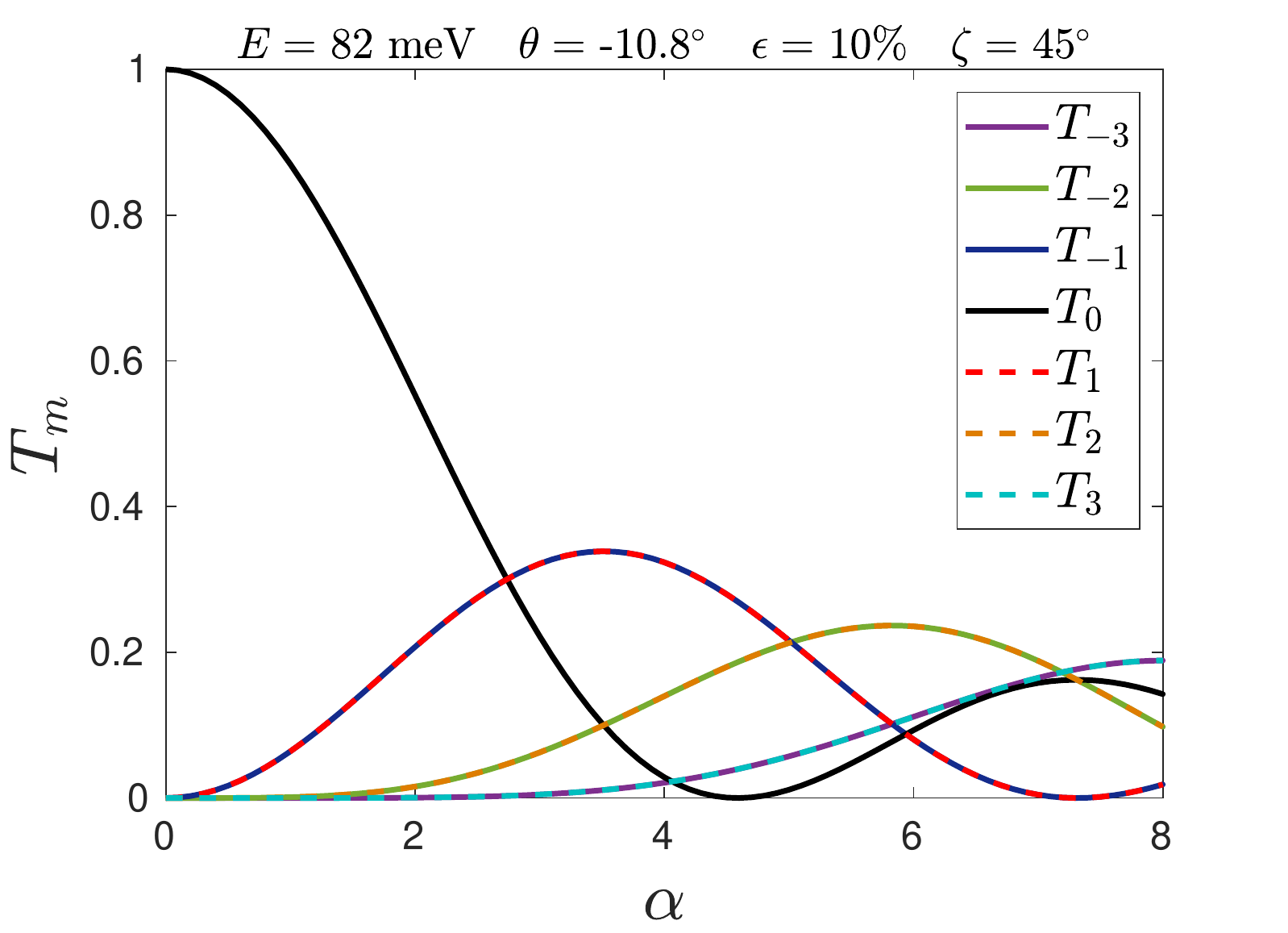} &
\includegraphics[trim = 1mm 0mm 12mm 0mm, scale= 0.39, clip]{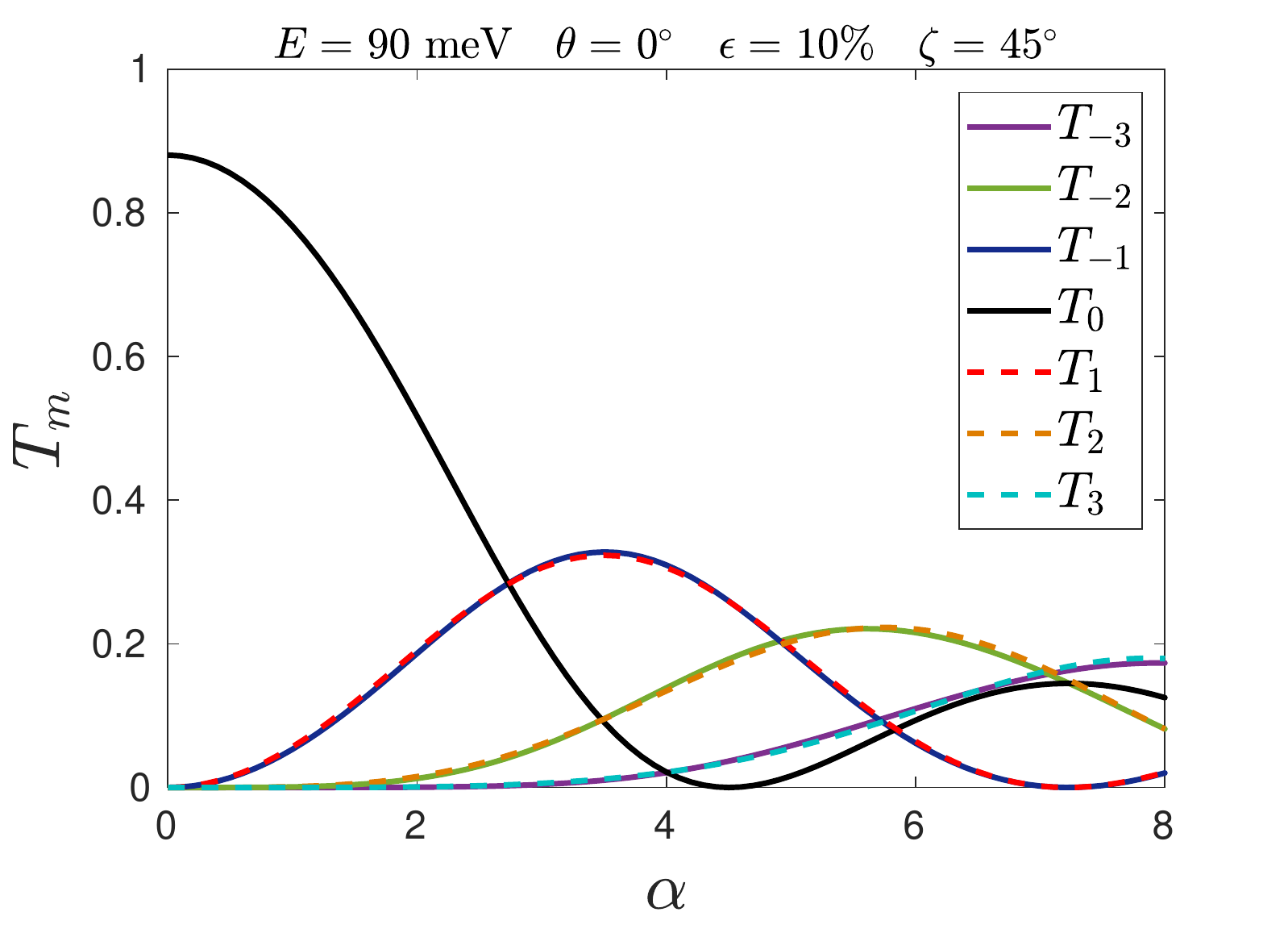}&
\includegraphics[trim = 1mm 0mm 12mm 0mm, scale= 0.39, clip]{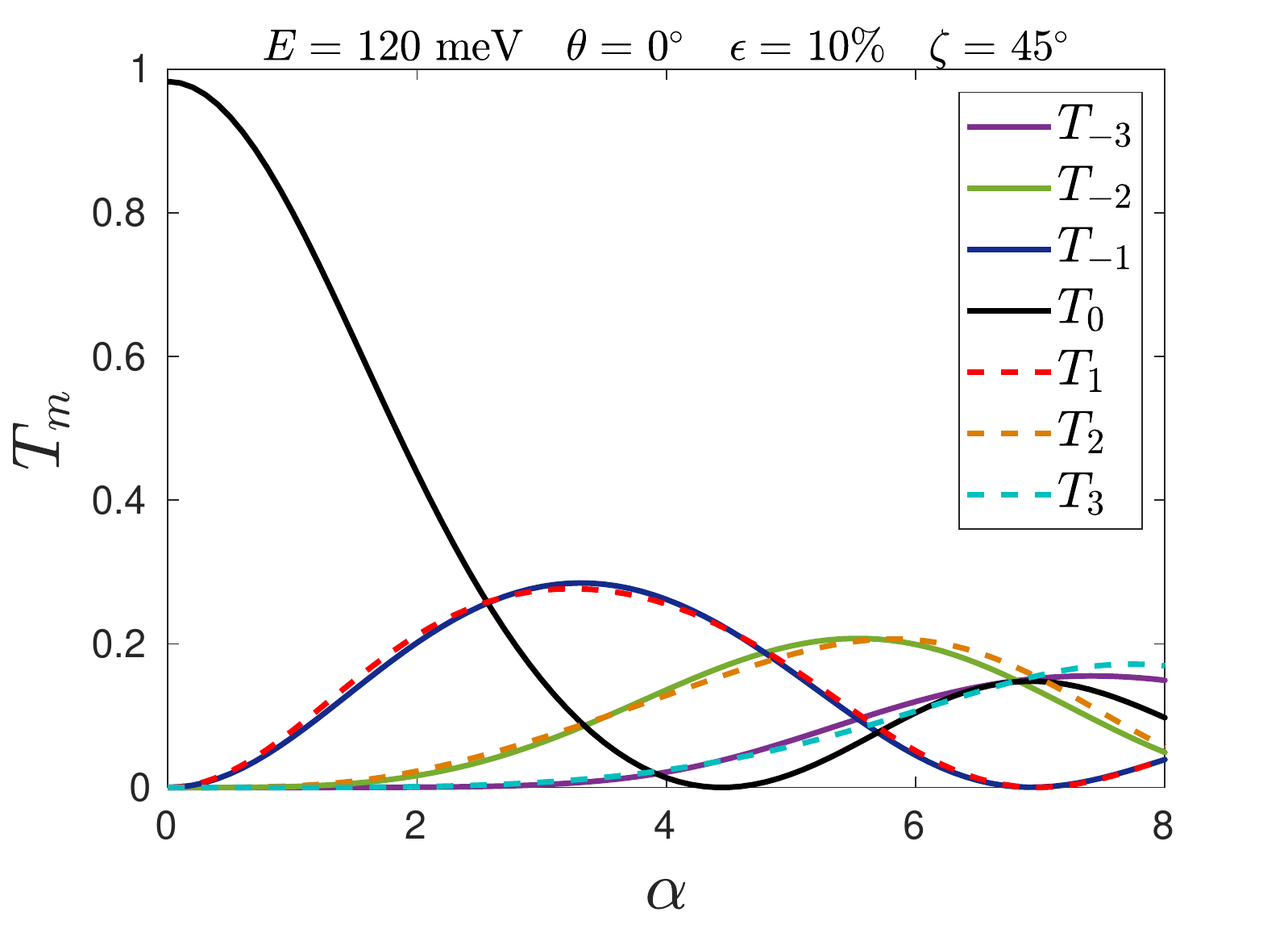}
\end{tabular}
\caption{Transmission probability $T_m = |t_m|^2$ for electrons with energy $E$ from the central band to cross the oscillating barrier and lie on the sideband $E - m\hbar\omega$. This transmission is obtained from the numerical solution of the linear equation 
system \eqref{M}, as a function of the ratio $\alpha = V_1/\hbar\omega$, where $V_1$ is the oscillation amplitude for a time-periodic potential of height $V_0 = 200$ meV, width $D = 100$ nm, and frequency $\omega = 5$ THz. The set of uniaxial strain parameters are $\epsilon = 10\%$ and $\zeta = 45\deg$. Transmission probability for the cases of anomalous Klein tunneling in the incidence angle $\theta_{KT} = -10.8\deg$ and using the energy $E = 82$ meV in (a), normal incidence in (b) and (c) for the energies $E = 90$ and $120$ meV, respectively.}
\label{fig2}
\end{figure*}
Now, in order to match the $e^{-im\omega t}$ behaviour in region II, we must introduce
the wave vectors $\vec k_m$ in regions I and III, defined by
\begin{subequations}
\begin{eqnarray}
\eee(s_m,\vec k_m^{\pm})&=&E-m\hbar\omega\\
k_{y,0}&=&k_{y,m}^{\pm}.
\end{eqnarray}
\label{eq:defk}
\end{subequations}
Note again that $\vec k_m^{\pm}$ are uniquely determined by $\vec k_0$. Also, the $\pm$ sign 
is related to a choice of left-going and out-going waves, see (\ref{kpr1}). Similarly, the solution of (\ref{eq:defk}) is given by:
\begin{subequations}
\begin{eqnarray}\label{kpr1}
k_{x,m} ^{\pm}& = & \pm s_m\sqrt{\frac{(E - m\hbar\omega)^2}{\hbar^2v_x^2}-{\frac{v^2_y}{v^2_x}k_{y,0} ^2\sin^2(\mu_x+\mu_y)}} \nonumber\\
& & -k_{y,0}\tan\theta_{KT},\\ 
s_m&=&\sgn(E-m\hbar\omega).
\end{eqnarray}
\end{subequations}

It is possible to express $k_{y,0}$ in terms of the incidence angle $\theta$ inverting Eq. \eqref{theta}
\begin{equation}\label{ky}
    k_{y,0} = \frac{v_x|E|(\tan\theta - \tan\theta_{KT})}{\hbar v^2_y\sin^2(\mu_x + \mu_y)
    \sqrt{1 + \frac{v^2_x(\tan\theta - \tan\theta_{KT})^2}{v^2_y\sin^2(\mu_x + \mu_y)}}}.
\end{equation}

We now make the following {\em Ansatz} for the wave function $\vec\psi(\vec r,t)$ in terms of the band index and wave vector values in the three different regions I, II, and III: 
\begin{subequations}
\begin{eqnarray}
 \vec{\psi}^I(\vec r,t) & = & \frac{1}{\sqrt{2}}e^{ik_{y,0}y}e^{-iE t/\hbar}\left[\binom{1}{s_0 e^{i \phi^+_0}}e^{i k^+_{x,0} x} \right.\nonumber\\
 &  & + \left.\sum_{m=-\infty}^\infty r_m\binom{1}{s_m e^{i\phi^-_m}}e^{i k^-_{x,m}x}e^{-im \omega t}\right], \nonumber\\
 &  &\\
 \vec{\psi}^{II}(\vec r,t) & = &\frac{1}{\sqrt{2}}e^{ik_{y,0}y}e^{-iE t/\hbar}\!\!\!\!\!\!\sum _{n,{m}=-\infty}^\infty  \!\!\!\!\!\!J_n(\alpha)\left[t'_{m}\binom{1}{s'_m e^{i\xi^+_m}}\right.\nonumber \\
 & & \!\!\!\!\!\! \times e^{iq^+_{x,m}x} + \left.r'_{m}\binom{1}{s'_me^{i\xi^-_m}}e^{iq^-_{x,m}x}\right]e^{-i {(n+{m})\omega}t},\nonumber\\
 &&\\
 \vec{\psi}^{III}(\vec r,t) & = &\frac{1}{\sqrt{2}}e^{ik_{y,0}y}e^{-iE t/\hbar}\!\!\!\!\sum_{m=-\infty}^\infty \!\!\!\!t_m\binom{1}{s_m e^{i \phi^+_m}}\nonumber \\
 && \qquad \qquad \times e^{i k^+_{x,m} x}e^{-im \omega t},
\end{eqnarray}
\label{eq:ansatz}
\end{subequations}

\noindent where we define the phases corresponding to the various wave vectors, as described in \eqref{defphi}

\begin{eqnarray}\label{phases}
\phi_m^{\pm}&=&\phi(\vec k_m^{\pm})\\
\xi_m^{\pm}&=&\phi(\vec q_m^{\pm})
\end{eqnarray}

\noindent and we use the particular eigenstates \eqref{eq:defphi}. The coefficient $r_m$ is the reflection amplitude of the incident wave back into region I, $t'_m$ and $r'_m$ are the amplitudes of the right-going and left-going waves in region II
respectively, and $t_m$ is the total transmission amplitude from
I to III, while gaining or losing an energy $m\hbar\omega$ in the process.  
The sideband index indicates the conduction ($s'_m = 1$) or valence ($s'_m = -1$) band.

With the matching of the wave functions (\ref{eq:ansatz}) at $(x = 0)$ 
and $(x = D)$ and using the orthonormality condition of Fourier basis, we obtain the following equations system

\begin{subequations}
\begin{eqnarray}
\delta_{m0} + r_m  &=& \!\!\! \sum_{l=-\infty}^{\infty}J_{m-l}(\alpha)(t'_l + r'_l)\\ \label{23}
s_m\delta_{m0} e^{i{\phi}_m^{+}}+ s_m r_m e^{i{\phi}_m^{-}} & = & 
 \!\!\!\sum_{l=-\infty}^{\infty}s'_l J_{m-l}(\alpha)\left(t'_l 
e^{i{\xi}_l^{+}} \right. \nonumber\\
&& \quad +\left.r'_l 
e^{i{\xi}_l^{-}}
\right)\\\label{24}
t_m
e^{i k_{x,m}^{+} D}&=&\!\!\!\sum_{l=-\infty}^{\infty}J_{m-l}({\alpha})
\left(t'_l 
e^{iq_{x,l}^{+}D}\right.\nonumber\\
&&\left.\quad +r'_l e^{iq_{x,l}^{-} D}
\right)\\\label{25}
s_mt_m e^{i k_{x,m}^{+}D}e^{i\phi_m^{+}} & = & \!\!\!\sum_{l=-\infty}^{\infty}s'_lJ_{m-l}({\alpha})
\left(
t'_l e^{iq_{x,l}^{+}D}e^{i\xi_l^{+}}\right.\nonumber\\
&&\left.\qquad + r'_l e^{iq_{x,l}^{-} D}e^{i\xi_l^{-}}
\right)\label{26}
\end{eqnarray}
\label{lineqs}
\end{subequations}

The linear equations system \eqref{lineqs} must be truncated up to a maximum number of terms because 
in principle, it is infinite. We can define this maximum number $L$ in the sum and impose the conditions
\begin{equation}
r_{m} = t'_{m} = r'_{m} = t_{m} = 0\qquad(|m|\geq L+1).
\end{equation}
In this way, the dimension of the system is $d \times d$, 
where $d = 4(2L + 1)$, and the sum index $m$ runs over $-L$ to $L$ in resemblance to the $L_z$ 
angular momentum quantization. We chose the ordered basis for the amplitude coefficients defining 
the vector of $d$ components by 

\begin{eqnarray}
\vec{X} &=& (r_{-L}, \ldots, r_{L},t'_{-L}, \ldots, t'_{L}, r'_{-L}, \ldots, r'_{L}, t_{-L}, \ldots, t_L)^{\textrm{T}}
\nonumber\\
&=&\left[
\big(r_m\big)_{m=-L}^L\big(t^\prime_m\big)_{m=-L}^L\big(r^\prime_m\big)_{m=-L}^L\big(t_m\big)_{m=-L}^L
\right]^{\textrm{T}}
\end{eqnarray}
We now write the equations system \eqref{lineqs} in a slightly more compact form as follows:
\begin{subequations}
\begin{eqnarray}
{\mathcal M}_1 &=& [-\mathcal{I} \mathcal{J} \mathcal{J} \mathcal{O}],\\
{\mathcal M}_2 &=& [\mathcal{O} \mathcal{J} \mathcal{J} -\mathcal{I}],\\
\mathcal{J} & = & {\mathcal J}_{ml} = J_{m -l}(\alpha),\\
{\mathcal D}_1&=& \diag\left[\left(s_me^{i\phi_m^{-}}\right)_{m=-L}^L \left(s'_m e^{i\xi_m^{+}}\right)_{m=-L}^L \right.\nonumber\\
&&\quad\ \left.\left(s'_me^{i\xi_m^{-}}\right)_{m=-L}^L \left(s_me^{i\phi_m^{+}}\right)_{m=-L}^L\right],\\
\mathcal{D}_2&=&
\diag\left[\left(e^{ik_{x,m}^{-}D}\right)_{m=-L}^L \left(e^{iq_{x,m}^{+}D}\right)_{m=-L}^L\right. \nonumber\\
&&\qquad\left.\left(e^{iq_{x,m}^{-}D}\right)_{m=-L}^L \left(e^{ik_{x,m}^{+}D}\right)_{m=-L}^L\right],\\
\vec b_1&=&\big(
\delta_{m,0}
\big)_{m=-L}^L\ ,\\
\vec{b}_2&=&s_0e^{i\phi_0} \vec b_1
\end{eqnarray}
\end{subequations}
where the square sub-matrices $\mathcal{I}$ 
and $\mathcal{O}$ are the identity and null matrix of size  $d/4 \times d/4$, respectively.
Therefore, equations (\ref{lineqs}) can be written as , 
\begin{subequations}
\begin{eqnarray}
\mathcal{M}_1\vec{X} &=& \vec{b}_1\\
\mathcal{M}_1\mathcal{D}_1\vec{X} &=& \vec{b}_2\\
\mathcal{M}_2\mathcal{D}_2\vec{X} &=& \vec{0}\\
\mathcal{M}_2\mathcal{D}_1\mathcal{D}_2\vec{X} &=& \vec{0}
\end{eqnarray}
\label{eq:defmat}
\end{subequations}

\begin{figure*}[t!!]
\begin{tabular}{ccc}
(a) \qquad \qquad \qquad \qquad \qquad \qquad \qquad \qquad& (b) \qquad \qquad \qquad \qquad \qquad \qquad 
\qquad & (c) \qquad \qquad \qquad \qquad \qquad \qquad \qquad\\
\includegraphics[trim = 1mm 0mm 12mm 0mm, scale= 0.41, clip]{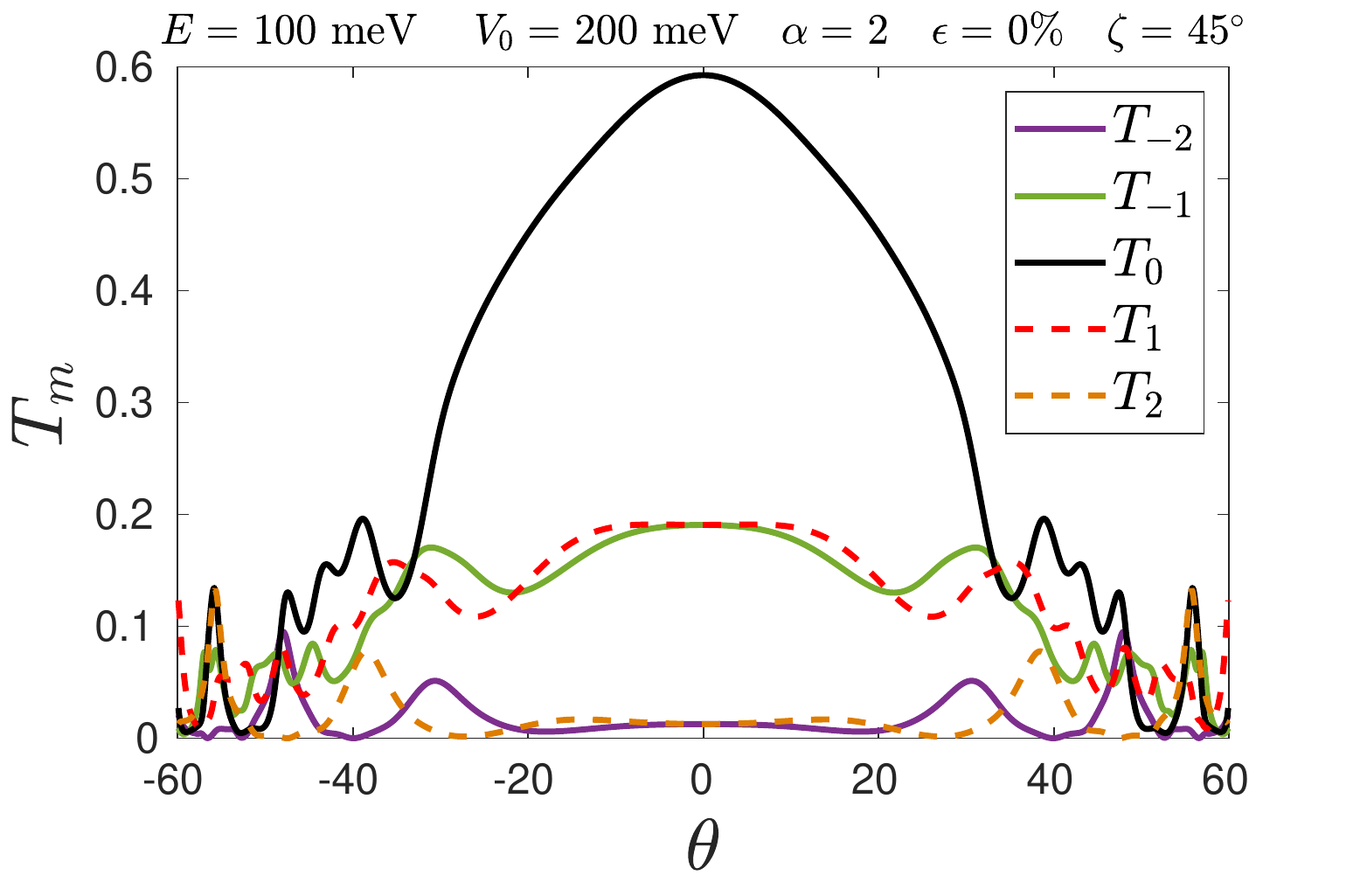} &
\includegraphics[trim = 1mm 0mm 12mm 0mm, scale= 0.41, clip]{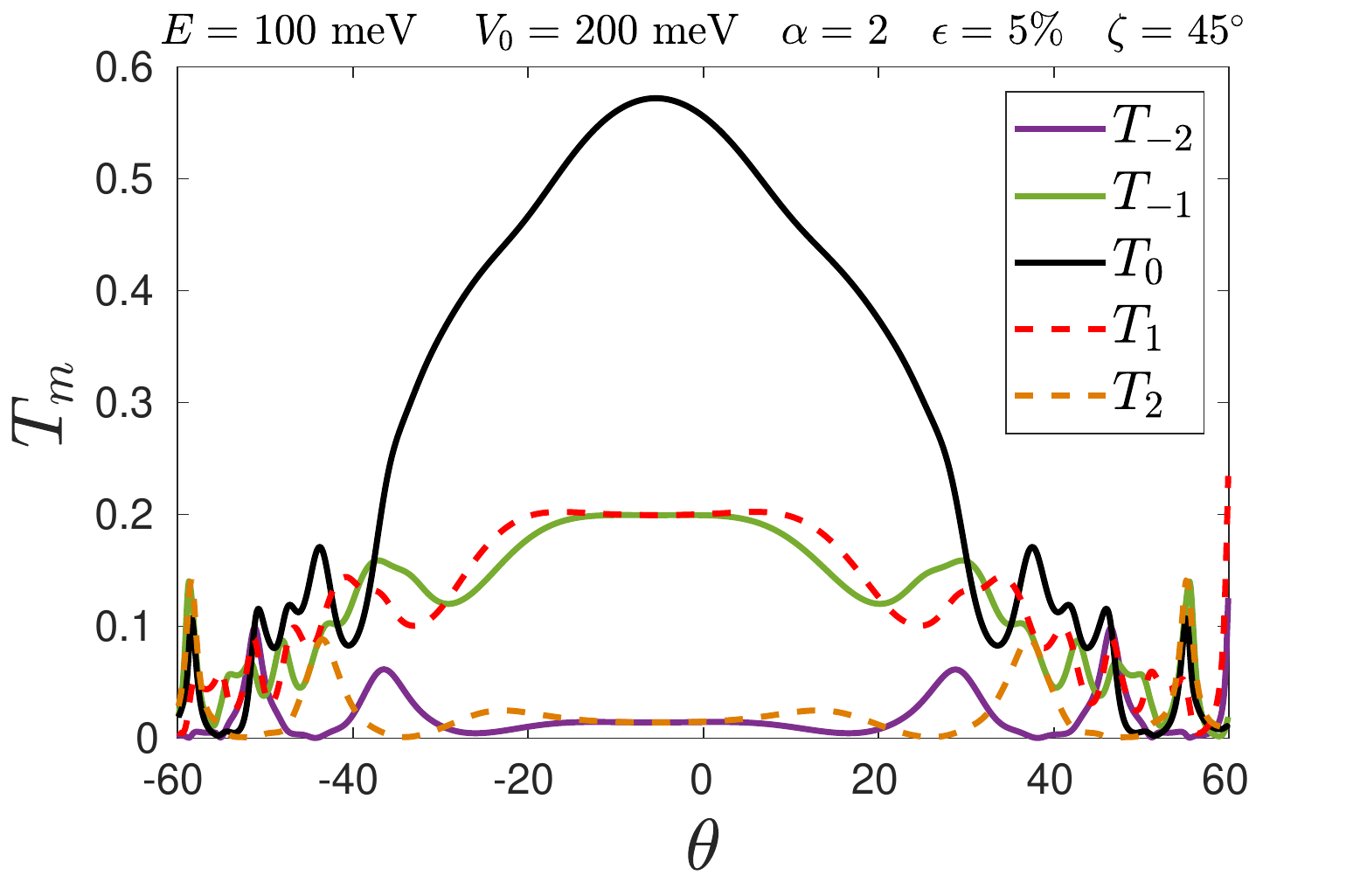} &
\includegraphics[trim = 1mm 0mm 12mm 0mm, scale= 0.41, clip]{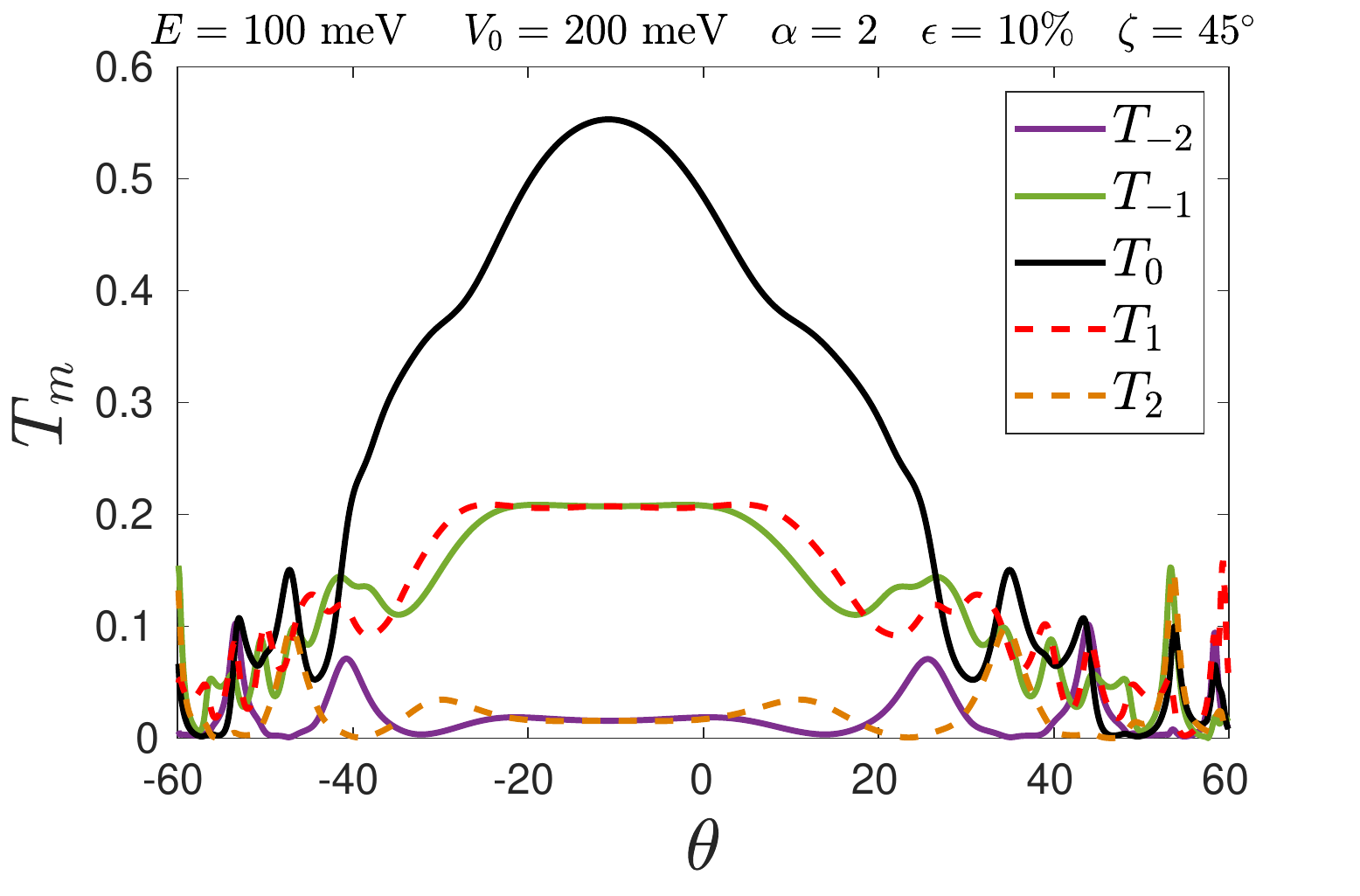}
\end{tabular}
\caption{Anomalous Floquet tunneling of electrons at the energy $E = 100$ meV 
as a function of the incidence angle $\theta$. The set of values for the time-periodic potential are $V_0 = 200$ meV, 
$D = 100$ nm, and $\omega = 5$ THz. Transmission probabilities $T_m = |t_m|^2$ using the strain parameters 
$\epsilon = 0, 5, 10\%$ and $\zeta = 45\deg$ with $\alpha = 2$ in (a), (b), and (c) respectively. }
\label{fig3}
\end{figure*}
\begin{figure}[t!!]
\includegraphics[trim = 0mm 0mm 0mm 0mm, scale= 0.53, clip]{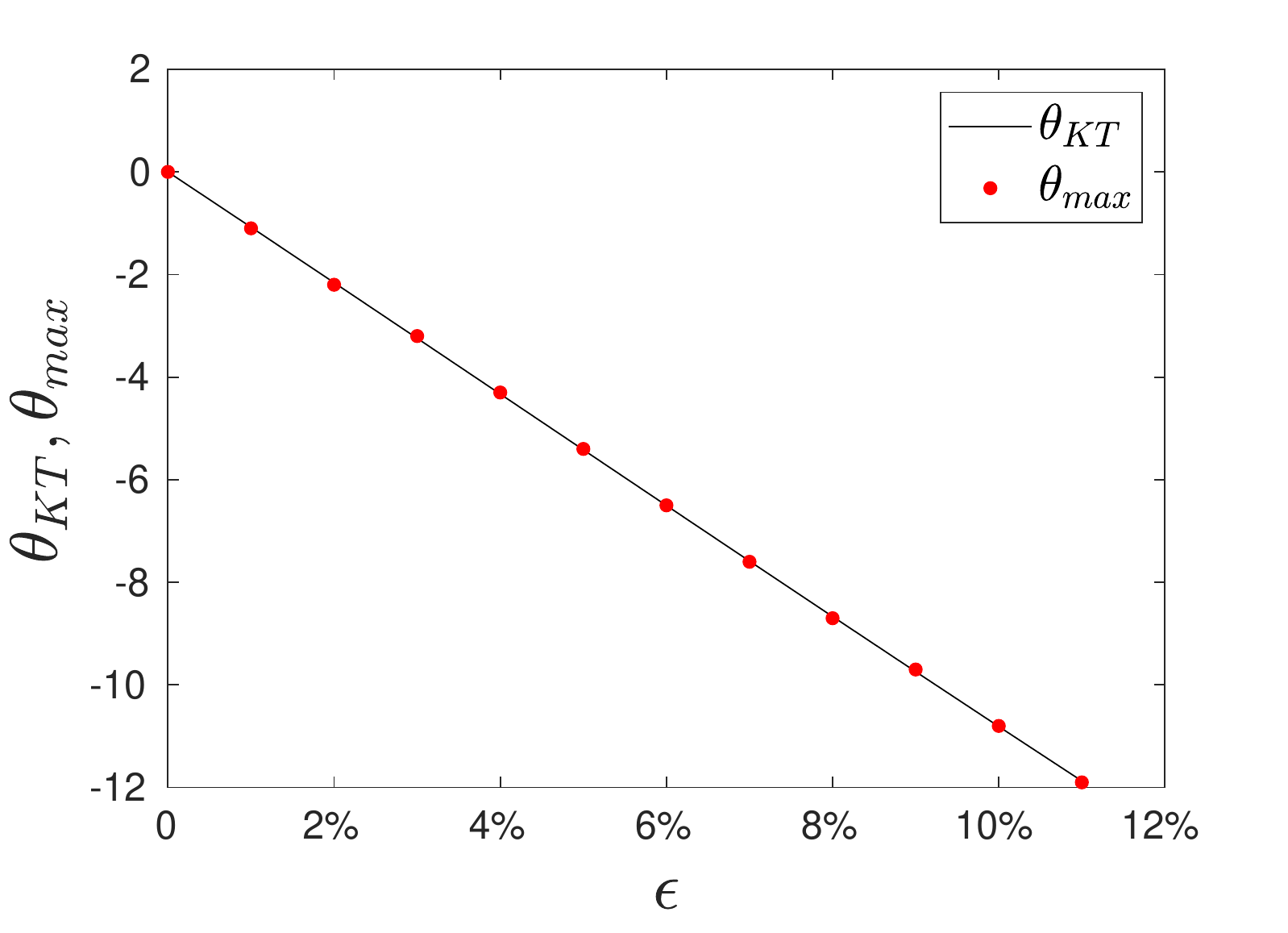}
\caption{Anomalous Klein tunneling angle (Eq. \eqref{KTA}) and the numerically determined angular shift of the maximum transmission $T_0$ as a function of the tensile strain parameter $\epsilon$ at the direction $\zeta = 45\deg$ and the ratio $\alpha = V_1/\hbar\omega = 2$.}
\label{ThKT}
\end{figure}

On the one hand, we can identify that the rectangular matrices $\mathcal{M}_1$ and $\mathcal{M}_2$ control the 
scattering of electrons in the time-periodic potential barrier through a unique tunable parameter $\alpha$. On the 
other hand, the diagonal matrices $\mathcal{D}_1$ and $\mathcal{D}_2$ contain the effect of strain from the 
phases in \eqref{phases} and wave vectors given by Eqs. \eqref{kpr} and \eqref{kpr1}. The photon-assisted tunneling amplitudes $t_m$ are provided by the last $2L + 1$ components of vector 
\begin{subequations}
\begin{eqnarray}
\vec{X} &=& \mathcal{M}^{-1}\vec{b},\\
\mathcal{M}&=& [\mathcal{M}_1 \quad \mathcal{M}_1\mathcal{D}_1 \quad \mathcal{M}_2\mathcal{D}_2 \quad \mathcal{M}_2\mathcal{D}_1\mathcal{D}_2]^{\textrm{T}}.
\label{M}
\end{eqnarray}
\end{subequations}
Here $\cal{M}$ is the total square matrix of the system defined by Eqs. \eqref{23}-\eqref{26} and 
$\vec{b} = (\vec{b}_1, \vec{b}_2, \vec{0}, \vec{0})$. Therefore, the coefficients are given by $T_m = |t_m|^2$ which quantify the transmission probabilities of electrons from the central band $E$ to cross the time-periodic potential barrier and transit to the sideband $E -m\hbar\omega$. 

 In appendix A, we show an approximate solution of this equation system with validity in the range $0 < \alpha < 1$.

\section{Discussion and results} \label{Discussion}

The application of uniaxial strain along the $\zeta = 45\deg$ changes drastically the transport properties in anisotropic
 hexagonal materials. Electrons impinging the electrostatic potential barrier at the specific incidence angle $\theta_{KT}$ 
 present the anomalous Klein tunneling \cite{Betancur-Ocampo2018}. This effect emerges for strains that break the 
 mirror symmetry with respect to the $x$ axis. We set the values $\epsilon = 10\%$ and $\zeta = 45\deg$, where 
 anomalous Klein tunneling appears for the incidence angle $ \theta_{KT} = -10.8\deg$ in the static barrier $\alpha = 0$, 
 (see Fig.~\ref{fig2}(a)). This perfect transmission occurs when the wave vector is perpendicular to the barrier, as 
 obtained setting $k_{y,0} = 0$ in Eq. \eqref{theta}. The incidence angle is different to zero due to that the wave 
 vector, pseudo-spin, and group velocities are generally not parallel \cite{Betancur-Ocampo2018, Betancur-Ocampo2019}. 
 If we turn on the time-periodic potential, the anomalous Klein tunneling suppresses. The transmission probabilities in the central and sidebands depend on $\alpha$. In most of the cases, we only consider the transmission coefficients $T_m$ with $m = -2,-1,0,1,2$  because the other ones with $|m| > 2$ have a maximum value smaller than 0.1 in the whole range of $0 < \alpha < 8$ 
 and therefore, they can be neglected. The transmission probability $T_m$ starts to be relevant for $\alpha > |m|$, as shown in Fig. \ref{fig2}. We can see that electrons absorbing or emitting $m$ photons have the same probability to 
 cross the barrier (see Fig. \ref{fig2}(a)). This equiprobability appears for the specific case where the wave vector is perpendicular to the barrier and also by the linear dispersion relation of electrons. 
 \begin{figure}[t!!]
  \begin{tabular}{c}
  \includegraphics[width=0.45\textwidth]{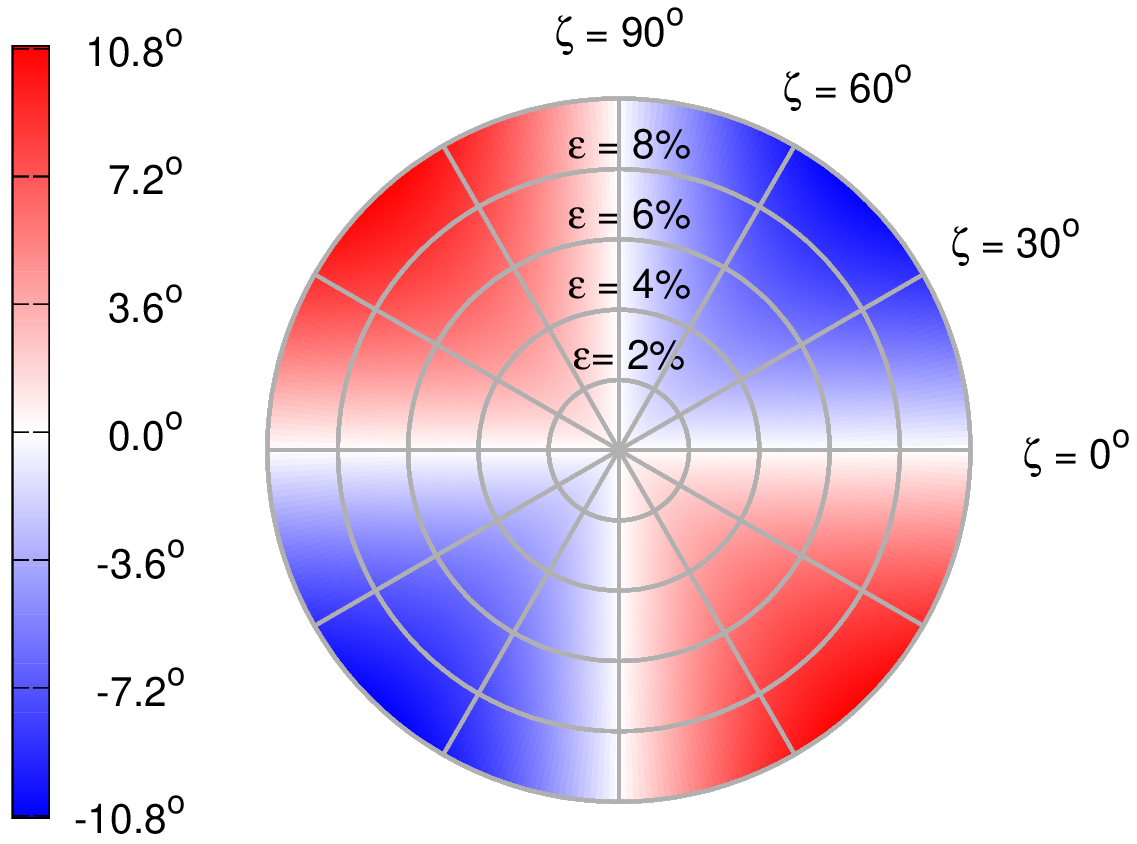}
  \end{tabular}
  \caption{Anomalous Klein tunneling angle as a function of the strain parameters $\epsilon$ (radius from 0 to $10 \%$) and $\zeta$ (polar angle in degrees).}
  \label{Anom}
\end{figure}
\begin{figure*}[t!!]
\begin{tabular}{ccc}
(a) \qquad \qquad \qquad \qquad  \qquad \qquad \qquad \qquad \qquad & (b) \qquad \qquad 
\qquad \qquad \qquad \qquad \qquad \qquad & (c) \qquad \qquad \qquad \qquad \qquad \qquad \qquad \qquad\\
\includegraphics[trim = 2mm 0mm 0mm 0mm, scale= 0.14, clip]{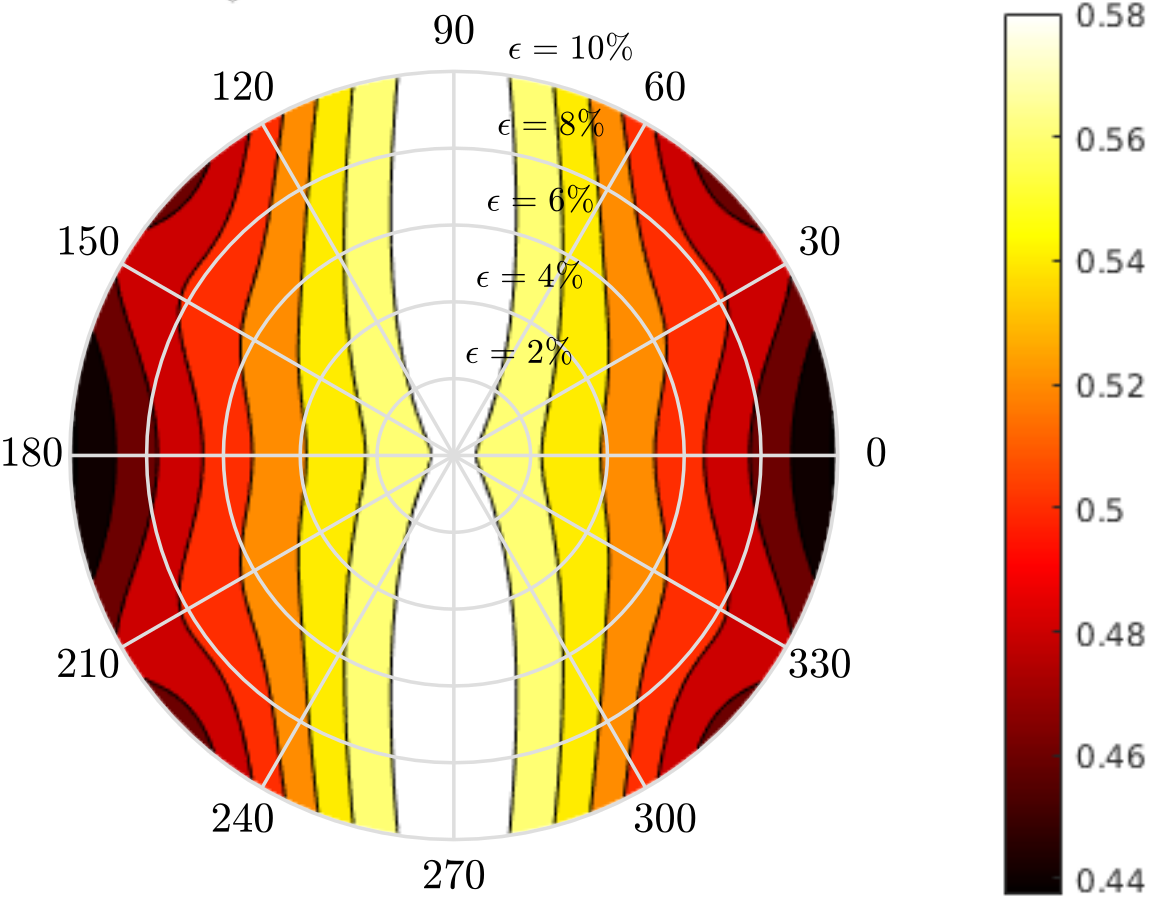} &
\includegraphics[trim = 2mm 0mm 0mm 0mm, scale= 0.14, clip]{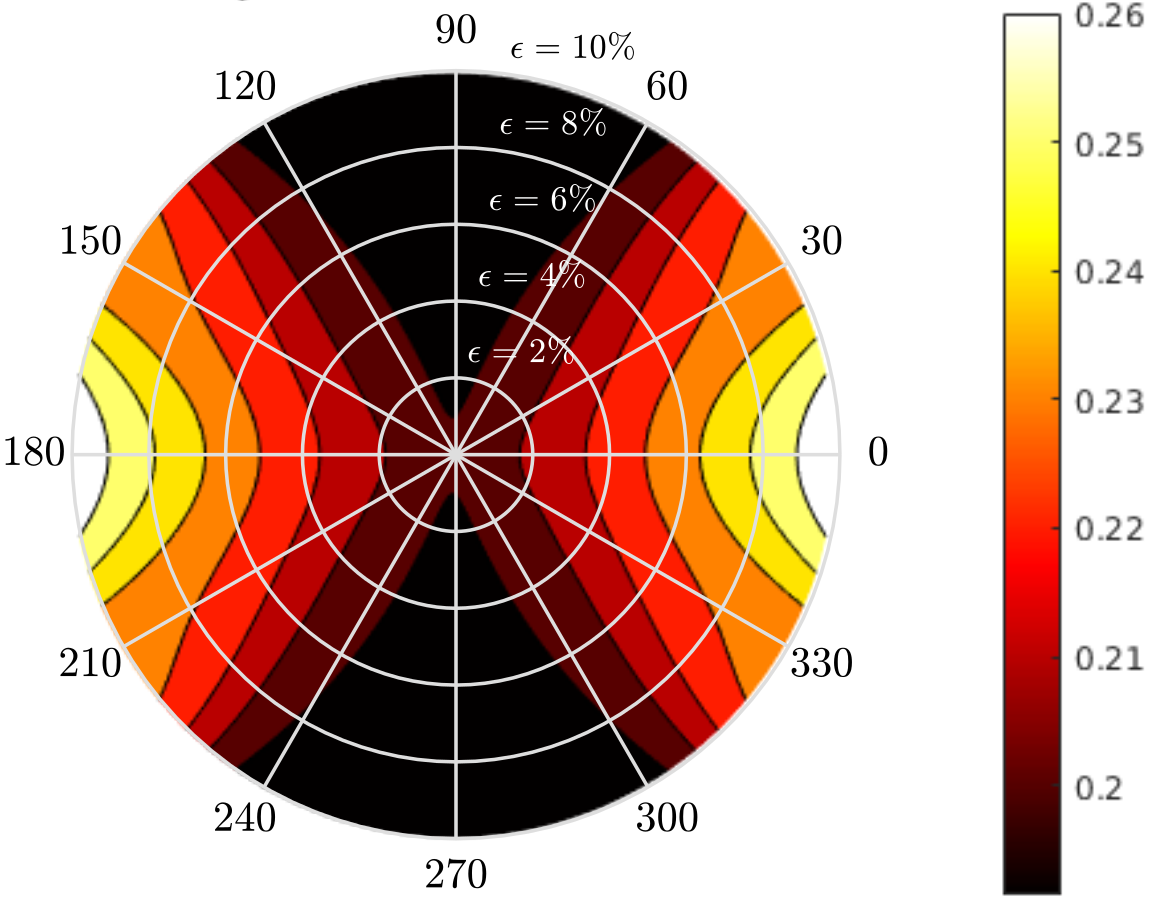}&
\includegraphics[trim = 2mm 0mm 0mm 0mm, scale= 0.14, clip]{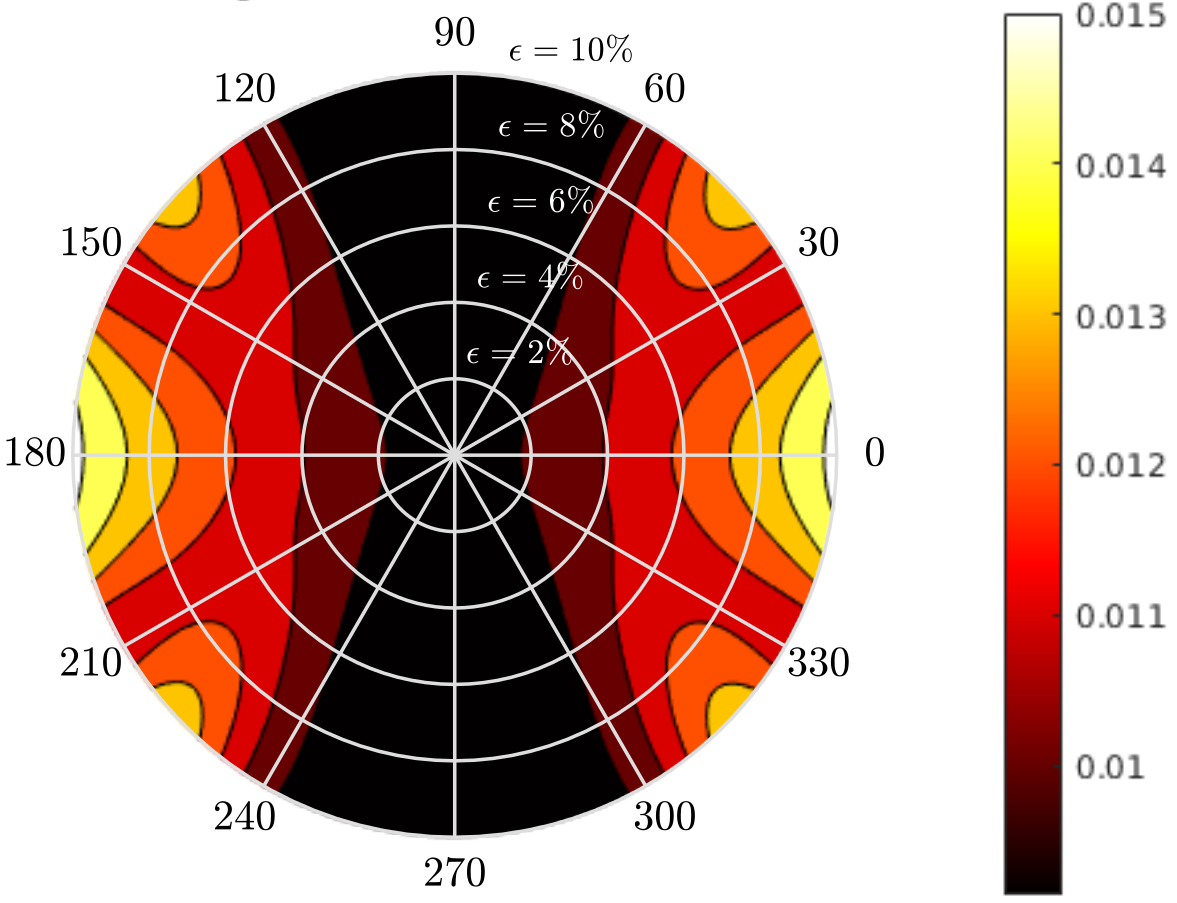}
\end{tabular}
\caption{Transmission probability (a) $T_0$, (b) $T_{-1}$, and (c) $T_{-2}$ as a function of uniaxial strain 
parameters $\epsilon$ (radius from 0 to $10 \%$) and $\zeta$ (polar angle in degrees) for electrons with 
normal incidence and energy $E = 100$ meV in a time-periodic potential $V_0 = 200$ meV, $D = 100$ nm, 
$\alpha = 2$, and $\omega = 5$ THz.}
\label{Tmvsstr}
\end{figure*}

 When electrons impinge under normal incidence to the time-periodic potential, as shown in Figs. \ref{fig2}(b) and (c), 
 the wave vector is not perpendicular anymore as a consequence of the uniaxial strain out of the main axes $x$ and $y$. Hence, Klein tunneling 
 deviates from the normal direction, and the transmission probability splits out slightly for the absorption and emission of 
 photons, namely, $T_m \neq T_{-m}$. The perfect transmission for normal incidence by the Klein tunneling in the static case $\alpha = 0$ is destroyed, as verified by changing the energy values to $E = 90$ and $120$ meV in Fig. \ref{fig2}(b) and (c). This resonant tunneling is atypical for normal incidence.
\begin{figure*}[t!!]
\begin{tabular}{ccc}
(a) \qquad \qquad \qquad \qquad  \qquad \qquad \qquad \qquad  & (b) \qquad 
\qquad \qquad \qquad \qquad \qquad \qquad & (c) \qquad \qquad \qquad \qquad \qquad \qquad \qquad\\
\includegraphics[trim = 0mm 0mm 12mm 0mm, scale= 0.4, clip]{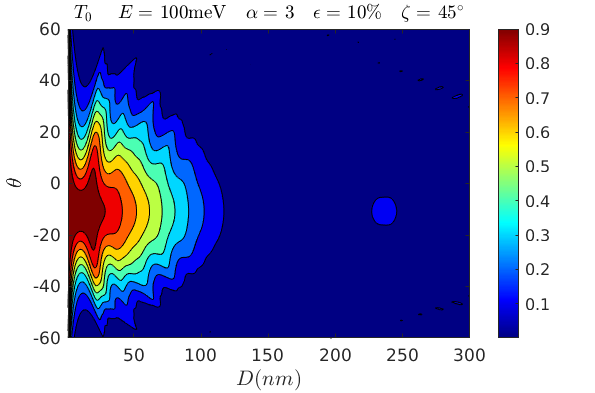}&
\includegraphics[trim = 0mm 0mm 12mm 0mm, scale= 0.4, clip]{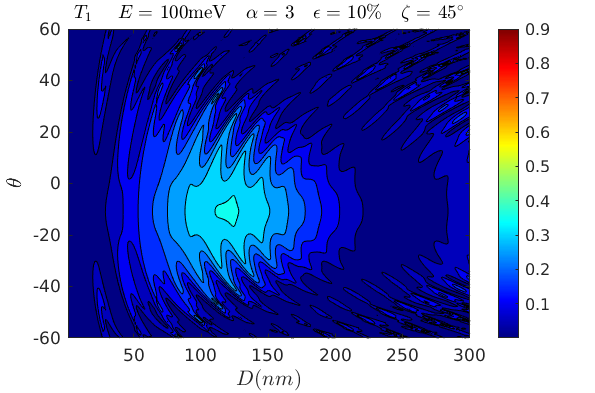}&
\includegraphics[trim = 0mm 0mm 12mm 0mm, scale= 0.4, clip]{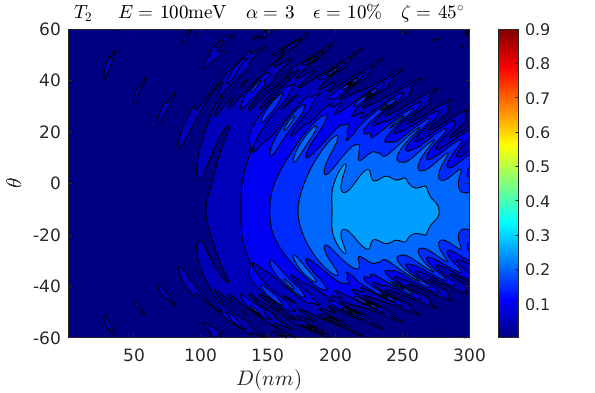}
\end{tabular}
\caption{Transmission probabilities of (a) $T_0$, (b) $T_1$, and (c) $T_2$ as a function of barrier width and incidence angle for the potential height $V_0 = 200$ meV.}
\label{Tmvsbarr}
\end{figure*}

We call anomalous Floquet tunneling to the angular shift of the maximum transmission by the application of a uniaxial tension different to the direction $\zeta = 0\deg$ and $90\deg$, as shown in Figs.~\ref{fig3} (a)-(c). Such a shift in the transmission probability usually appears in other systems by time-reversal symmetry breaking with an external magnetic field \cite{Sinha2012}. That vector potential, which 
generates the magnetic field, shifts the Dirac cone in the reciprocal space. Although the strain here affects the 
Dirac cone differently, changing the circular shape to a rotated and elliptical one, both systems present a 
particular feature in common: incident electrons with a wave vector perpendicular to the interface have a nonzero 
parallel group velocity $v_y$. In Fig.~\ref{fig3}, we chose a shortened incidence angle range to avoid the 
evanescent waves. The incident electrons have critical angles that depend on the sideband. Increasing 
$\epsilon$ in Figs. \ref{fig3}(a)-(c), we observe that this angular deviation 
in the maximum of transmissions improves. We show the angular shift of the maximum transmission $\theta_{max}$ as a function of the tensile strain in Fig. \ref{ThKT}. This angular shift has a good agreement with the anomalous Klein tunneling angle $\theta_{KT}$ predicted by Eq. \eqref{KTA}. It is worth noting that this angular shift of the maximum transmission depends only on the tensile strain $\epsilon$ and tension angle $\zeta$. The expansion of Eq. \eqref{KTA} (see appendix B), keeping only the first-order terms in $\epsilon$, we lead to a very simple and straightforward relation of the anomalous Klein tunneling angle with the parameters $\epsilon$ and $\zeta$
\begin{equation}\label{linthetaKT}
    \theta_{KT} \approx \frac{360\deg}{\pi}\rho^+(1-\beta)\epsilon\sin2\zeta,
\end{equation}
\noindent which has a negligible deviation of the exact relation \eqref{KTA} in the whole strain range considered. Uniaxial strain along with the directions $\zeta = 0\deg$ and $90\deg$ (not shown) does not break the mirror symmetry with respect to the normal axis. In this case, the group velocity and the wave vector are parallel for normal incidence 
which restores the angular transmission symmetry. We quantify the anomaly in the transmissions using the direction of the Klein tunneling deviation in Eq. \eqref{KTA}, which depends only on the strain parameters $\epsilon$ and $\zeta$. Fig. \ref{Anom} shows this anomaly in the whole strain range. As expected, the uniaxial strains along the perpendicular and parallel directions to the interface keep the symmetry in the transmission. While tensions in a different direction to $\zeta \neq 0$ and $90\deg$ cause the anomalous Floquet tunneling. We found that the highest angular shift value is $\theta_{KT} \approx -10.8\deg$ for the set of parameters $\epsilon = 10\%$ and $\zeta = 45\deg$.
\begin{figure*}[t!!]
\begin{tabular}{ccc}
(a) \qquad \qquad \qquad  \qquad \qquad \qquad \qquad \qquad & (b) \qquad 
\qquad \qquad \qquad \qquad \qquad \qquad & (c) \qquad \qquad \qquad \qquad \qquad \qquad \qquad\\
\includegraphics[trim = 0mm 0mm 12mm 0mm, scale= 0.4, clip]{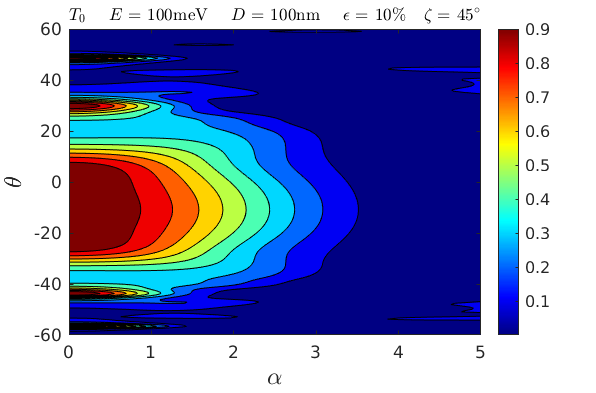}&
\includegraphics[trim = 0mm 0mm 12mm 0mm, scale= 0.4, clip]{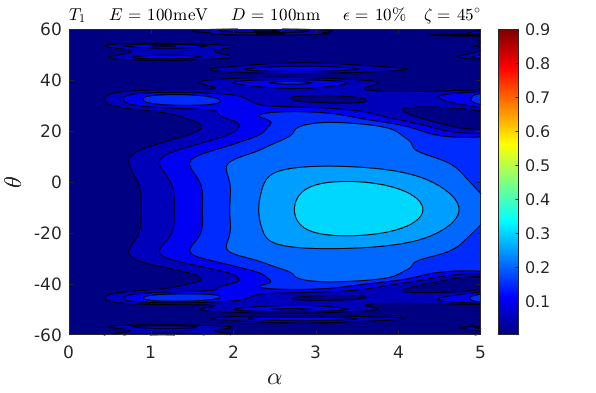}&
\includegraphics[trim = 0mm 0mm 12mm 0mm, scale= 0.4, clip]{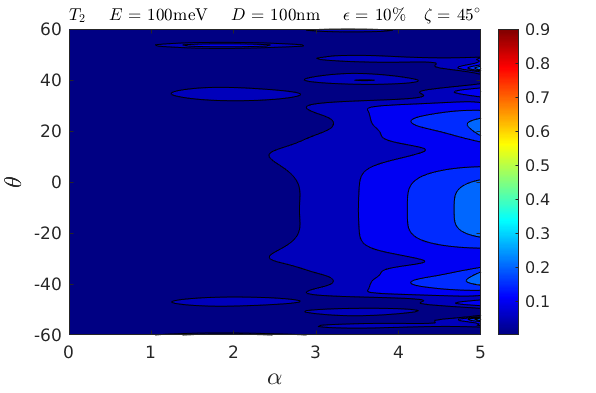}
\end{tabular}
  \caption{Transmission probabilities of (a) $T_0$, (b) $T_{1}$, and (c) $T_{2}$ as a function 
  of the parameter $\alpha$ and incidence angle $\theta$ for the potential height $V_0 = 200$ meV 
  and width $D = 100$ nm.}
\label{Tmvsalp}
\end{figure*}

To understand how the uniaxial strain affects the behavior of photon-assisted tunneling, we show the probability 
transmission as a function of $\epsilon$ and $\zeta$ in Fig.~\ref{Tmvsstr} for the case of normal incidence. We can see that in a wide range of $\zeta$, the behavior of $T_m$ is strongly anisotropic with the angle $\zeta$, and the increase of $\epsilon$ causes a reduction in the probability transmission $T_0$, as shown in Fig.~\ref{Tmvsstr}(a). However, for the tension angle $\zeta = 90\deg$, normal 
incident electrons have an almost constant probability of crossing the barrier regardless of the tensile strain. It is worth 
noting that the independence of transmission on the tensile strain at $\zeta = 90\deg$ also appears for other sidebands, as shown in Fig. \ref{Tmvsstr}(b) and (c). The transmission $T_{1}$ and $T_{2}$ present an identical behavior with respect to the emission counterpart. The application of strain in the directions near the $x$-axis shows an increase of $T_{-1}$ and $T_{-2}$. While uniaxial strain along the $y$-axis decrease the electron transmission in the sidebands. The time exposition of electrons to the oscillating barrier explains the strain-induced transition from elastic to inelastic tunneling. Positive tensile strains in the direction $\zeta = 0\deg$ increase the bond lengths. Therefore, the probability amplitude of electrons decreases to hop among neighboring sites. In this way, there is more time exposition to interact with the time-periodic potential. Thus, electrons cross the barrier inelastically with transmission probabilities $T_1$ and $T_2$. In contrast, deformations parallel to the interface decrease the zigzag bond lengths, and electrons have a major probability to cross the barrier elastically.

We examine the behavior of the transmission probability $T_m$ as a function of barrier width and incidence 
angle, as shown in Fig.~\ref{Tmvsbarr}. In general, the reminiscence of the anomalous Klein tunneling makes that 
almost all the transmission occurs around the incident angle $\theta = -10.8\deg$. We found that the tuning of barrier 
width can serve as a selector of the transmission in the sidebands. In thin barriers $D < 50$ nm (see 
Fig.~\ref{Tmvsbarr}(a)), the transmission is mainly due to the central band, where other sidebands participate only scarcely. The increase of the barrier width can suppress the transmission in the central band 
and favors the emergence of another transmissions in the sidebands. Fig.~\ref{Tmvsbarr}(b) shows that electrons 
absorbing or emitting one photon have a higher probability of crossing the time-periodic barrier if the width 
is within the range of $100$ to $150$ nm. The same occurs for $T_2$ in Fig.~\ref{Tmvsbarr}(c) in the range 
$150 < D < 300$ nm. This is due that electrons to cross the barrier have more exposition time to interact 
with the time-periodic potential, and therefore, it favors the promotion of electrons to travel through other sidebands with a higher energy difference.  We note a similar behavior (not shown) for the 
transmissions $T_{-1}$ and $T_{-2}$ compared with $T_1$ and $T_2$, respectively. These results imply 
that an adequate selection of the barrier width allows that the device converts incoming electron 
current with energy $E$ to two outcoming photo-excited currents, with a difference between 
them of $2n\hbar\omega$. Although this effect can also be obtained for the unstrained case \cite{Freitag2012, Tielrooij2013, Ma2018}, the uniaxial deformation improves the inelastic tunneling to favor the output of photon-excited currents.

Another alternative way to select transmission in a particular sideband is to modulate the oscillating amplitude $V_1$. Fig.~\ref{Tmvsalp} shows the transmissions $T_0$, $T_{1}$, and $T_{2}$ as a function of $\alpha$ 
and $\theta$ for a constant value of the barrier width. Anomalous Klein tunneling and resonant peaks 
are suppressed by increasing $\alpha$, while electron transmissions in other sidebands arise. Dependent on the 
amplitude of the oscillation, the device in Fig.~\ref{fig1}(a), can convert electron current to a 
photo-excited one. 

\section{Conclusions and final remarks}\label{Conclusions}

We have studied the effect of uniaxial strain on the transport properties of electrons in graphene in the presence of a 
photon-assisted tunneling mechanics. The interplay of uniaxial strain and photon-assisted tunneling opens possibilities to control electron flow. We applied the Floquet scattering theory in anisotropic hexagonal lattices.
 This approach serves to understand the interaction of electron current with the time-periodic potential in systems 
 such as uniaxially strained graphene, photonic crystals, molecular graphene, and optical lattices. We calculate the 
 transmission probabilities with the absorption or emission of multiphoton processes. The main transmission 
 features as anomalous Floquet tunneling occur with the application of uniaxial strains out of the $x$ and $y$ axes. We found that applying uniaxial strain in the parallel direction at the interface, photon-assisted tunneling is unaffected by the increase of the tensile parameter. Whereas, uniaxial strain perpendicular to the barrier enhances the electron transmission from the sidebands. An appropiate design of the barrier width, or tuning the amplitude of oscillation, can select the electron tunneling to absorb or emit $n$ photons. 
 Therefore, the device converts an electron current to a photo-excited one. Such findings may be useful to control 
 the electron flow in nanoelectronic devices through the photon-assisted tunneling and strain engineering.

\section*{Acknowledgments}
PM ~gratefully acknowledges a fellowship from UNAM-DGAPA, YBO and TS from CONACYT Project A1-S-13469, CONACYT Project Fronteras 952, and the UNAM-PAPIIT research grant IA-103020. FL and DE~acknowledge financial support from CONACYT Project 254515. We thank T.H. Seligman and L.E.F. Foa-Torres for useful discussions and comments. 
 
\appendix
\section*{Appendix A: Approximate solution of Floquet scattering of electrons in uniaxially strained graphene}
\setcounter{section}{1}
\setcounter{equation}{0}
\label{app:apprsol}

It is possible to obtain an approximate solution for the transmission coefficient $T_j = |t_j|^2$ with $j = -1$ and $1$ 
using the exposed method in section \ref{Floquet}. As we can see, the fact that the $J_{n}({\alpha})$ is negligible at $n \gg L$ in the 
range $0 < \alpha < L$, it causes that the infinite system evolves a finite one from $-L$ up to $L$. In the case 
$L = 0$, the equation system has dimension $d = 4$ and we can calculate the transmission coefficient for the static barrier
 \begin{widetext}
\begin{eqnarray}\label{Tsn}
T_{sm} =\frac{\cos^2\chi_m\cos^2\chi'_m}{\cos^2\chi_m\cos^2\chi'_m\cos^2\gamma'_m + [1 - s_m s'_m\sin\chi_m\sin\chi'_m]^2\sin^2\gamma'_m},
\end{eqnarray}
\end{widetext}
\noindent which is the probability of an electron to cross the barrier from the same sideband with energy $E - m\hbar\omega$, where 
\begin{subequations}
\begin{eqnarray}
\chi_m &=& \phi_m^{+} + \mu_x,\\
\chi'_m&=& \xi_m^{+} + \mu_x,\\
\gamma'_m  &=&\frac{D |E - V_0 -m \hbar\omega|}{v_x \hbar}\cos(\xi_m^{+} + \mu_x).
\end{eqnarray}
\end{subequations}
 Due to the dependence on barrier width in Eq. \eqref{Tsn}, resonant tunneling occurs for $\gamma'_m = N\pi$ being $N$ an integer. While 
 anomalous Klein tunneling appears for the incidence angle $\theta = \theta_{KT}$. Without deformation, the above values are 
 $\mu_x =0$, $\mu_y = \pi/2$, $v_x = v_y = 1$, recovering the expression of transmission coefficient in a static 
 barrier of graphene \cite{Katsnelson}. With the definition of transmission probability in photon-assisted tunneling 
 $T_m=|t_m|^2$ and solving the equation systems for $L = 1$, we find an analytical transmission for the transmission $T_{1}$ valid in the range $0 < \alpha < 1$ 

\begin{equation}\label{T1}
T_1 = \left(\frac{J_1(\alpha)|(\Gamma_1 - \Gamma_0) + (\Lambda_1 - \Lambda_0)|}
{J_0(\alpha)
\left|
e^{i\phi_1^{+}} -
e^{i\phi_1^{-}}
\right|}\right)^2T_{s0}T_{s1},
\end{equation}

\noindent where the quantities $\Gamma_j$ and $\Lambda_j$ are defined as
\begin{subequations}
\begin{eqnarray}
    \Gamma_j&=& \frac
    {
    e^{iq_{x,j}^{-}D}
    (e^{i\phi_1^{-}} + e^{i\xi_j^{-}})
    (e^{i\phi_0^{+}} + e^{i\xi_j^{+}})
    }{
    e^{i\xi_j^{-}} - e^{i\xi_j^{+}}
    }
\\
    \Lambda_j &=& 
    \frac
    {
    e^{i q_{x,j}^{+}D}(e^{i\xi_j^{+}} +e^{i\phi_1^{-}})(e^{i\phi_0^{+}} +
     e^{i\xi_j^{-}})
     }
     {e^{i\xi_j^{-}} - e^{i\xi_j^{+}}
     },
\end{eqnarray}
\end{subequations}
with the index $j = 0$ or $1$. An identical expression is obtained for the transmission 
$T_{-1}$ replacing $1 \rightarrow -1$ in all the relations above. In the unstrained case, the transmission probability \eqref{T1} 
is identical to those calculated in \cite{Zeb2008}.

\section*{Appendix B: Linear relation of anomalous Klein tunneling angle with the uniaxial strain}
\setcounter{section}{2}
\setcounter{equation}{0}

In order to obtain the linear dependence on the tensile strain $\epsilon$ of anomalous Klein tunneling angle, we expand Eq. \eqref{KTA} keeping the first-order in $\epsilon$. First, we calculate the ratio of the complex velocities components \eqref{Complvs}
\begin{equation}\label{ComplKT}
    \frac{v^c_y}{v^c_x} = \frac{a_{1y}\tau_1{e}^{-i\vec{K}_D\cdot\vec{\delta}_1}
 + a_{2y}\tau_2{e}^{-i\vec{K}_D\cdot\vec{\delta}_2}}{a_{1x}\tau_1{e}^{-i\vec{K}_D\cdot\vec{\delta}_1}
 + a_{2x}\tau_2{e}^{-i\vec{K}_D\cdot\vec{\delta}_2}}.
\end{equation}
\noindent This expression is useful to express Eq. \eqref{KTA} as
\begin{widetext}
\begin{equation}\label{KTAappr}
     \theta_{KT} \approx \frac{180\deg}{\pi}\text{Re}(v^c_y/v^c_x)= \frac{180\deg}{\pi}\frac{a_{1x}a_{1y}\tau^2_1 + a_{2x}a_{2y}\tau^2_2 + (a_{1x}a_{2y} + a_{2x}a_{1y})\tau_1\tau_2\cos[\vec{K}_D\cdot(\vec{\delta}_1-\vec{\delta}_2)]}{a_{1x}^2\tau^2_1 + a_{2x}^2\tau^2_2 + 2a_{1x}a_{2x}\tau_1\tau_2\cos[\vec{K}_D\cdot(\vec{\delta}_1-\vec{\delta}_2)]}.
\end{equation}
\end{widetext}

Taking into account that the solution for Eq. \eqref{DPs} is

\begin{equation}\label{DPsol}
    \cos[\vec{K}_D\cdot(\vec{\delta}_1 - \vec{\delta}_2)] = \frac{\tau^2_3 - \tau^2_2 - \tau^2_1}{2\tau_1\tau_2}
\end{equation}

\noindent and the deformed lenghts of uniaxially strained graphene are

\begin{equation}
    \delta_j \approx a\{1 + \rho^-\epsilon + \rho^+\epsilon\cos[2\zeta + (2j - 1)60\deg]\},
\end{equation}

\noindent we expand the exponential decay rule for the hopping parameters $\tau_j$ up to first-order in $\epsilon$

\begin{equation}\label{tauappr}
    \frac{\tau_j}{\tau} \approx 1 - \beta\{\rho^-+\rho^+\cos[2\zeta + (2j - 1)60\deg]\}\epsilon.
\end{equation}

\noindent Substituting the above expression in \eqref{DPsol}

\begin{equation}\label{ApprDPsol}
    \cos[\vec{K}_D\cdot(\vec{\delta}_1 - \vec{\delta}_2)] \approx -\frac{1}{2}[1 + 3\beta\rho^+\epsilon\cos(2\zeta - 60\deg)].
\end{equation}
\noindent In the same way, we expand the relations

\begin{eqnarray}
\label{a1x}
a_{1x}\tau_1 &\approx& \sqrt{3}a\tau(1 + c_{1x}\epsilon),\\
a_{1y}\tau_1 &\approx& \sqrt{3}a\tau\rho^+\epsilon\sin2\zeta,\\
a_{2x}\tau_2 &\approx& \frac{\sqrt{3}}{2}a\tau(1 + c_{2x}\epsilon),\\
\label{a2y} a_{2y}\tau_2 &\approx& \frac{\sqrt{3}}{2}a\tau(\sqrt{3} + c_{2y}\epsilon),
\end{eqnarray}

\noindent where we used Eqs. \eqref{lattvs} and \eqref{tauappr}. The coefficients $c_{1x}$, $c_{2x}$, and $c_{2y}$ are, respectively,

\begin{eqnarray}
c_{1x} & = & \rho^- + \rho^+\cos2\zeta - \beta[\rho^- + \rho^+\cos(2\zeta+60\deg)],\\
c_{2x} & = & \rho^- + 2\rho^+\cos(2\zeta-60\deg) - \beta[\rho^- - \rho^+\cos2\zeta],\\
c_{2y} & = & \sqrt{3}\rho^- + 2\rho^+\sin(2\zeta-60\deg) - \sqrt{3}\beta[\rho^- - \rho^+\cos2\zeta].\nonumber\\
&&
\end{eqnarray}

Substituting the relations \eqref{a1x}-\eqref{a2y} and \eqref{ApprDPsol} on Eq. \eqref{KTAappr}, we obtain

\begin{eqnarray}
    \theta_{KT} &\approx& \frac{180\deg}{\pi}\frac{2\rho^+(1-\beta)\epsilon\sin2\zeta}{1 + 3[c_{1x} + c_{2x} -\beta\rho^+\cos(2\zeta - 60\deg)]\epsilon} \nonumber\\
    &\approx& \frac{360\deg}{\pi}\rho^+(1-\beta)\epsilon\sin2\zeta,
\end{eqnarray}

\noindent which is the result as shown in Eq. \eqref{linthetaKT}.


\end{document}